\documentclass{aa}  

\usepackage{natbib}
\usepackage{txfonts}
\usepackage{hyperref} 
\usepackage{mathrsfs}
\usepackage{graphicx}
\usepackage{ulem}

\newcommand{\Hunits}{km s$^{-1}$ Mpc$^{-1}$ }

\begin{document} 
\title{On the ellipticity parameterization for an NFW profile: an overlooked angular structure in strong lens modeling}

\author{Matthew R. Gomer\inst{1}\fnmsep \thanks{\email{mgomer@uliege.be}}
         \and
          Dominique Sluse\inst{1}
        \and
          Lyne Van de Vyvere\inst{1}
        \and
          Simon Birrer\inst{2,3,4}
        \and 
          Anowar J.~Shajib\inst{5, 6, 7}
        \and
          Frederic Courbin\inst{8}
        }

  \institute{STAR Institute, Quartier Agora - All\'ee du six Ao\^ut, 19c B-4000 Li\`ege, Belgium 
  \and 
  Department of Physics and Astronomy, Stony Brook University, Stony Brook, NY 11794, USA
  \and
  Kavli Institute for Particle Astrophysics and Cosmology and Department of Physics, Stanford University, Stanford, CA 94305, USA
  \and
  SLAC National Accelerator Laboratory, Menlo Park, CA, 94025
  \and
  Department  of  Astronomy  \&  Astrophysics,  University  of Chicago, Chicago, IL 60637, USA
    \and
    Kavli Institute for Cosmological Physics, University of Chicago, Chicago, IL 60637, USA
    \and NHFP Einstein Fellow \and
  Institute of Physics, Laboratory of Astrophysics, Ecole Polytechnique F\'ed\'erale de Lausanne (EPFL), Observatoire de Sauverny, 1290 Versoix, Switzerland
             }

  \date{Received date; accepted date}

 
  \abstract
    {Galaxy-scale gravitational lenses are often modeled with two-component mass profiles where one component represents the stellar mass and the second is an NFW profile representing the dark matter. Outside of the spherical case, the NFW profile is costly to implement, and so it is approximated via two different methods; ellipticity can be introduced via the lensing potential (NFWp) or via the mass by approximating the NFW profile as a sum of analytical profiles (NFWm). While the NFWp method has been the default for lensing applications, it gives a different prescription of the azimuthal structure, which we show introduces ubiquitous gradients in ellipticity and boxiness in the mass distribution rather than having a constant elliptical shape. Because unmodeled azimuthal structure has been shown to be able to bias lens model results, we explore the degree to which this introduced azimuthal structure can affect the model accuracy. We construct input profiles using composite models using both the NFWp and NFWm methods and fit these mocks with a power-law elliptical mass distribution (PEMD) model with external shear. As a measure of the accuracy of the recovered lensing potential, we calculate the value of the Hubble parameter $H_0$ one would determine from the lensing fit. We find that the fits to the NFWp input return $H_0$ values which are systematically biased by about $3\%$ lower than the NFWm counterparts. We explore whether such an effect is attributable to the mass sheet transformation (MST) by using an MST-independent quantity, $\xi_2$. We show that, as expected, the NFWm mocks are degenerate with PEMD through an MST. For the NFWp,  an additional bias is found beyond the MST due to azimuthal structures {\it exterior to the Einstein radius}. 
    We recommend modelers use an NFWm prescription in the future, such that azimuthal structure can be introduced explicitly rather than implicitly.}

    \titlerunning{Azimuthal structure in elliptical NFW models}
   \maketitle
%

\section{Introduction}
Gravitational lensing allows for a direct measure of the mass of galaxies, whether or not that mass is visible, and as such is a valuable tool to study galaxies and dark matter. Lensing can also be a tool to study cosmology, because time delays between multiple images can be compared to those predicted from a lens model to constrain a time-delay distance $D_{\Delta t}\propto1/H_0$. These tools are only as accurate as the galaxy mass distribution models they depend on \citep[see reviews by e.g.,][]{Birrer22,Shajib22b}.

Galaxy lenses are generally massive elliptical galaxies which are often described using multiple mass components, with one tracing the light representing the baryon mass distribution and the other representing the dark matter mass distribution. Dark matter mass distributions are typically described using an Navarro Frenk White (NFW) profile \citep{Navarro96}, which takes the following form, parameterized in 3D in terms of a characteristic density $\rho_s$ and scale radius $r_s$:
\begin{equation}\label{eq:nfw_prof}
    \rho(r)=\frac{\rho_s}{(r/r_s)(1+r/r_s)^2}.
\end{equation} 
While the spherical 3D profile has a clean analytical form, the 2D-projected elliptical NFW profile, which is necessary for lensing applications, does not have an analytical form for the deflection angle and lensing potential. To circumvent this problem, \citet{Golse02} showed that a general elliptical lensing mass density profile can be expressed analytically if the ellipticity is introduced in the lensing potential rather than the mass distribution itself. This allows for analytical calculations of lensing properties, and so this parameterization has been the established method to model NFW profiles for various applications, including time-delay cosmography \citep[e.g.,][]{Wong20,Rusu20,Shajib22b}. Alternative methods to introduce ellipticity in the surface mass density while keeping the computation feasible include precalculating the numerical integrals on a grid to be interpolated \citep{Schramm90, Schneider91, Keeton01}, or expanding the mass profile into a series of analytically tractable components, such as the multi-Gaussian expansion \citep[MGE;][]{vandeVen10,Shajib19} or the sum of cored steep ellipsoids \citep[CSE;][]{Oguri21}. Because these methods introduce ellipticity directly in the mass distribution, they are thought to be more physically realistic. In this work we will compare the \cite{Golse02} representation of lens galaxies that captures the ellipticity in the lens potential, to the CSE-based method of \citet{Oguri21}, which  has the ellipticity in the mass-density. We will refer to these descriptions as the NFWp and NFWm parameterizations, respectively.

This work aims at quantifying the impact of these different parametrizations on the predicted lensing properties, with a specific interest on galaxy-scale time delay cosmography. The main feature of interest is that while both parameterizations have the same radial profile (defined as the 1D profile of convergence within circular annuli), the azimuthal structure is different because of the different ways in which ellipticity is introduced. \citet{Kochanek21} showed the azimuthal freedom of a lens model must be considered with great care, which encourages us to be wary of the exact prescription of azimuthal structure within lens models. Furthermore, \citet{VandeVyvere22a,VandeVyvere22b} showed in individual lenses how unmodeled azimuthal structure can bias lens models, even though as a population these effects appear to average out. Our concern is that the choice of NFW parameterization could represent a systematic bias in the treatment of azimuthal structure. As DM is a significant mass component in galaxies and as it is in many cases represented with a NFW profile, we focus here on comparing two different parametrizations of this profile and to quantify their impact on lens models. To quantify this comparison, we use the $H_0$ value recovered from the lensing model, as it directly reflects the capacity of a model to capture differences in the Fermat potential at the location of the images.


While the azimuthal structure of interest comes from the NFW prescription, we test the effect of this prescription in the case where the profile has two components where only the dark matter component is represented by an NFW profile, such that the test is more applicable to the analysis of observed systems. A sensible method to evaluate the impact of the choice of parametrization of the NFW on $H_0$ consists in emulating and modeling strongly lensed systems that resemble known ones. \citet{Gomer22} (hereafter TDCVIII) created a population of mock lens images from analytical profiles designed to match the observed population of TDCOSMO lenses. The profiles used to create these lenses were two-component profiles, with a Chameleon profile \citep{Dutton14} representing the light and an NFWp profile representing the dark matter. TDCVIII then fit these systems with a Power Law Elliptical Mass Distribution \citep[PEMD;][]{Barkana98} model with external shear. In this work, we create mocks which are identical to those used by TDCVIII, except that we change the NFW parameterization to create a systematic comparison between the NFWp and NFWm implementations. We fit these mocks in the same manner as TDCVIII and compare the results of the parameter inferences.

We quantify the impact of these parameterizations in the context of the Mass Sheet Transformation (MST), detailed in Section \ref{ssec:xiformalism}, where lenses with different radial profile shapes can give the same observables (i.e. image positions and fluxes). We make use of the MST-independent quantity $\xi_2$ (defined in Sect.~\ref{ssec:xiformalism}) to diagnose the degree to which the degeneracies in this paper can be attributed to the MST. The primary goal of this work is to evaluate the degree to which the choice NFW parameterization can play a role in the determination of $H_0$ and identify the source of that role in the context of lensing degeneracies.

The paper is structured as follows: Section \ref{sec:formalism} reviews the NFW profile parameterizations as well as $\xi$, Section \ref{sec:setup} compares the mock populations and the results of the PEMD fits, Section \ref{sec:discussion} discusses these results, Section \ref{sec:otherworks} expands this discussion to its implications for other works, and Section \ref{sec:conclusion} summarizes and concludes this work. Appendix \ref{sec:ellipticity_matching} gives an analytical description of the axis ratio of the mass of the NFWp parameterization and Appendix \ref{sec:appendix_systematics} details several subtleties regarding the calculation or $\xi$ in this work.
This work adopts a fiducial flat $\Lambda$CDM cosmology with $H_0=70$ \Hunits and $\Omega_{\rm M}=0.3$.

\section{Formalism}
\label{sec:formalism}

In this section, we review the key differences between the NFWp and NFWm parameterizations of the density profile (Sect.~\ref{ssec:NFW}), and discuss the quantity $\xi$ which we propose to use as a diagnostic of degeneracies between models beyond the Mass Sheet Transformation (Sect.~\ref{ssec:xiformalism}). 

\subsection{NFW parameterization}
\label{ssec:NFW}
Mass distributions of gravitational lenses are described in terms of surface mass density profiles normalized by the critical lensing density (i.e. convergence) $\kappa(r) = \Sigma(r) / \Sigma_{\rm crit}$. If one wishes to use the NFW profile for lensing applications, one must be able to accommodate ellipticity and project into the 2D plane of the sky. The convergence for a circular NFW profile can be expressed as \citep{Bartelmann96}
\begin{equation}\label{eq:nfw_kappa}
\kappa_{\rm NFW}=  \frac{2 \kappa_0}{u^2-1}[1-F(u)],
\end{equation}
with scaled radius $u=r/r_s$ and normalization $\kappa_0=\rho_s r_s/\Sigma_{\rm crit}$, and where
\begin{equation*}\label{eq:F(u)}
F(u)= \begin{cases} 
      \frac{1}{\sqrt{1-u^2}}{\rm arctanh}\sqrt{1-u^2} & (u<1)\\
      \frac{1}{\sqrt{u^2-1}}{\rm arctan}\sqrt{u^2-1} & (u>1).\\
      \end{cases}
\end{equation*}
The corresponding lensing potential can then be expressed as \citep{Meneghetti03}
\begin{equation}\label{eq:nfw_circ_pot}
\psi_{\rm NFW}=  2\kappa_0 \theta_s^2 h(u),
\end{equation}
where $\theta_s$ is the scale radius expressed in arcseconds and
\begin{equation*}\label{eq:h(u)}
h(u)= \begin{cases} 
      {\rm ln}^2\frac{u}{2} - {\rm arccosh}^2\frac{1}{u} & (u<1)\\
      {\rm ln}^2\frac{u}{2} + {\rm arccos}^2\frac{1}{u} & (u>1).\\
      \end{cases}
\end{equation*}

Ellipticity can be added to a given 1D profile by replacing the $r$ argument with an elliptical radius such as $r_{\rm ell}=\sqrt{q x^2+y^2/q}$, where $q$ is the axis ratio of the ellipse. For some profiles such as the Chameleon profile, this ellipticity is added directly to the convergence. However, such a substitution is not guaranteed to result in an analytical form for the lensing potential: the NFW profile has no such analytical form for the lensing potential arising from an elliptical 2D-projected mass distribution, instead requiring expensive numerical integrals to approximate.


The NFWp parameterization solves this problem by instead adding ellipticity analytically into the lensing potential, a strategy which can be applied generically for any analytical lensing potential profile. \citet{Golse02} show that one can add ellipticity to the potential by replacing $r$ with $r_{\epsilon}=r_{\rm ell} \sqrt{2q_\psi/(1+q_\psi^2)}$ in Eq. \ref{eq:nfw_circ_pot}, giving analytical expressions for the convergence and shear. To minimize confusion between ellipticity conventions, we have introduced $q_\psi$ as the axis ratio of the potential. Similarly, we will use $q_\kappa$ for the convergence, and reserve $q$ to a general context. Because of the analytical form of the lensing potential, the lensing calculation is fast to compute. For low ellipticity values, the mass distribution contours have an approximately elliptical shape. However, for high ellipticity values this leads to the mass distribution taking on nonphysical dumbbell-shaped contours, and so modelers often restrict use of the NFWp profile to low values of ellipticity, although the exact definition of "low" may differ depending on the specific lensing application and is ultimately a choice on the part of the modeler. This parameterization is the conventional approach to model NFW profiles in most lensing contexts. 

Alternatively, the NFWm parameterization, in which the ellipticity is directly implemented into the mass distribution of the NFW profile, is composed of a combination of cored steep ellipsoid (CSE) profiles with a joint centroid position, each of which has a core radius $S_i$ and amplitude $A_i$. Each CSE has the following convergence profile:
\begin{equation}
    \kappa_{\rm CSE} = \frac{A_i \kappa_0}{2(S_i^2+r_{\rm ell}^2)^{3/2}},
\end{equation}
where $r_{\rm ell}$ is used as the radial element.
The CSE profile has ellipticity introduced in the mass distribution, and has analytical forms for the lensing potential and its derivatives \citep{Keeton98}. \citet{Oguri21} provided an approximation of the NFW profile using 44 CSE components such that the sum emulates the NFW profile to a precision of $\sim 0.01\%$ in terms of the lensing potential, deflection, and convergence. Calculation of lensing quantities takes approximately ten times as long as the NFWp, but with the advantage that the elliptical mass contour shape is retained for all values of ellipticity.

The input ellipticity for the NFWp profile describes the potential and as such does not equate to the ellipticity of the mass distribution. Therefore, in order to make an apt comparison between the two parameterizations, we modify the input ellipticity to the NFWp profile such that the two parameterizations have the same mass ellipticity at small radii. In Appendix \ref{sec:ellipticity_matching}, we describe analytically the relationship between the axis ratio of the potential and that of the mass for a general profile, and detail our process to match these ellipticities in the specific case of the NFWp profile.

\begin{figure*}
    \centering 
    \includegraphics[width=0.95\linewidth]{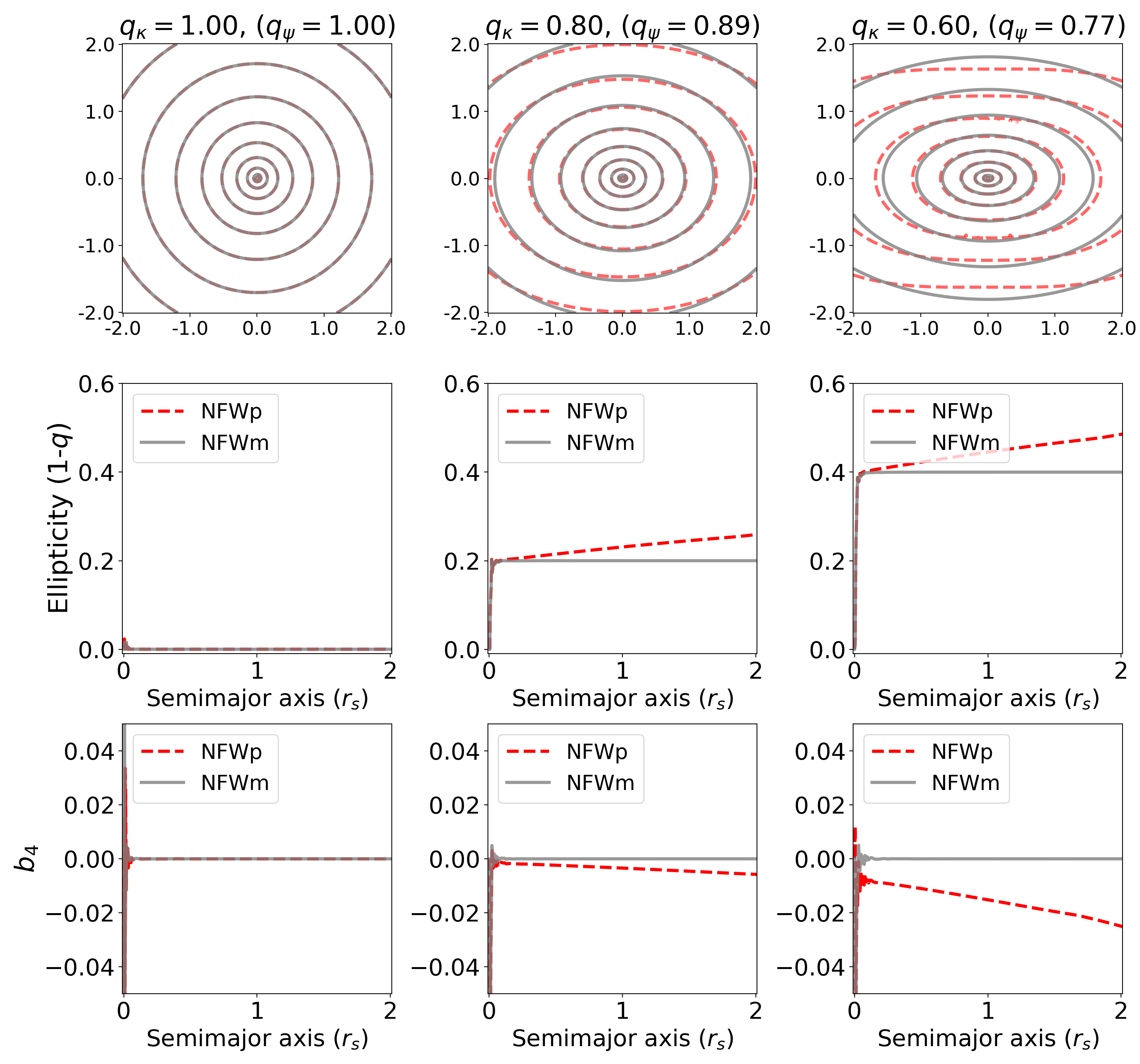}
    \caption{NFW 2D mass isodensity contours of $\kappa_{\mathrm{NFW}}$ (top), ellipticity as a function of semimajor axis (middle), and the $b_4$ multipole component as a function of semimajor axis (bottom), in units of NFW scale radius. The NFWp parameterization introduces ellipticity through the lensing potential and is shown in red while the CSE-based 
    NFWm parameterization introduces ellipticity directly in the mass and is shown in gray. The panels from left to right indicate the comparison for three input axis ratios. To match the $\kappa_{\mathrm{NFW}}$ ellipticities, we use the procedure in Appendix \ref{sec:ellipticity_matching}, which results in the NFWp potential having an input $q_\psi$ as indicated above the top panels.  }
    \label{fig:nfw_contours}
\end{figure*}

After matching the mass ellipticities, we show in Figure \ref{fig:nfw_contours} the shape of the mass distribution for an NFW profile using both the NFWp (red) and NFWm (gray) parameterizations of ellipticity. We show three different values of $q_\kappa$ (and corresponding $q_\psi$ in the NFWp case) as an input. In the circular case, the two profiles agree. In the elliptical case, the two profiles are not identically shaped but both appear physical. As the ellipticity increases, the NFWp contours become more oval-shaped, and if we were to further increase the ellipticity, they would become a nonphysical dumbbell shape \citep{Kassiola93}. This effect was also discussed in detail by \citet{Golse02}, and as such the NFWp profile is typically restricted in use to low values of ellipticity.

To quantify the azimuthal structure as a function of semimajor axis, we use the \texttt{photutils} software, which uses the method of \citet{Jedrzejewski87} to fit the elliptical structure of isophotes. The method iteratively fits the position angle and ellipticity of an ellipse. Once this fit is found, departures from ellipticity in terms of Fourier multipole components are calculated as
\begin{equation}
    I = I_0 + a_n \sin(n E) + b_n \cos(n E),
\end{equation}
where $n=3$ or $4$, $I$ is the intensity of a given contour, $I_0$ is the intensity of the fit ellipse contour, and $E$ is the eccentric anomaly. After a fit is achieved, the semimajor axis of the ellipse in question is increased and the process is repeated, such that the result is a description of the ellipticity, position angle, and Fourier amplitudes as a function of semimajor axis. We apply this fitting procedure to the isodensity contours and measure the ellipticity of the mass (expressed as $1-q$) as a function of semimajor axis, which we plot in the middle panels. In terms of the Fourier multipole components, the only component which returns nonzero values is $b_4$, a parameter of particular interest because nonzero $b_4$ is often measured in isophotes of real early type galaxies and is known to have an effect on $H_0$ recovery in lens systems \citep{VandeVyvere22a}. A positive value of $b_4$ indicates a "disky" shape and a negative value indicates a "boxy" shape. In the bottom panels, we plot the fit values of $b_4$, which indicate that the contours systematically become more boxy as a function of semimajor axis.


Setting aside the known $q$ mismatch effect (which we have accounted for) and the "dumbbell-shaped" effect (which does not apply in this ellipticity regime), in this work we identify a third difference in behavior between the two approaches; namely, there is a significant increase in ellipticity with radius. We refer to this effect as an "ellipticity gradient" in this work. Additionally, the mass distributions resulting from NFWp profiles present a nonzero boxiness which increases with radius, even in the case where the mass contours have physically realistic convex shapes. The presence of this type of azimuthal structure even in the low-ellipticity regime (i.e., before the dumbbell shapes arise) is somewhat overlooked in the literature, as in most use cases of the NFWp profile the intention is to mock an elliptical shape, and so an approximately constant elliptical shape is implicitly assumed in the low-ellipticity regime. 



\subsection{Mass Sheet Transformation and $H_0$}\label{ssec:xiformalism}
It is possible for two different mass distributions to reproduce the same imaging information. The most well-studied of such degeneracies is the Mass Sheet Transformation \citep[MST,][]{Falco85}, where the convergence is rescaled by a constant factor of $\lambda$ and a uniform sheet of mass\footnote{
        This transformation is purely mathematical in nature. This means that this sheet of mass does not have to be a real component missed by the model. Instead, it is possible that the distribution of the "true" mass profile of the lens and the chosen model are, to a good precision, mapped to each other via a MST \citep[e.g.][]{Schneider2013}.}
is correspondingly added:
\begin{equation}\label{eq:mst}
    \kappa_{\lambda}(r) = \lambda \kappa(r) + (1-\lambda).
\end{equation}
In addition, the unobservable source position is rescaled by the same factor,
\begin{equation} \label{eq:mst_beta}
\beta_{\lambda} = \lambda \beta.
\end{equation}
Under the MST, image positions and relative fluxes are unchanged. Meanwhile, time delays are affected by a factor of $\lambda$.

The MST is critical for lensing in a cosmological context, because the true mass distribution of the lens is unknown and lensing cannot distinguish between two mass profiles which are within an MST of one another. Because a given lens could be fit with more than one possible model (e.g., PEMD or composite), it is useful to describe lens profiles in a MST-independent way, meaning to relate only the quantities which lensing directly constrains rather than those which come from model choice. One such MST-independent quantity is the Einstein radius, $R_{\rm E}$, for which in this work we adopt the definition of the circular aperture within which the mean integrated surface mass density is equal to the critical density for lensing. The MST-independent nature of this quantity makes it one of the few quantities which lensing directly measures rather than infers from a model, and so it would be exceedingly useful to find more analogous quantities. The dimensionless quantity $\xi$, expressible using a combination of derivatives of the lensing potential $\psi$, has been developed for this purpose by \citet{Sonnenfeld17, Kochanek20, Birrer21b}. Together with the Einstein radius, $R_{\rm E}$, $\xi$ is an MST-invariant quantity, and so this work uses it as a metric to evaluate the degree to which two mass distributions are within an MST mapping from one to another.

This work primarily uses the \citet{Kochanek20} $\xi_2$ definition, which is defined by Taylor expanding deflection for a circular lens for an image near $R_{\rm E}$:
\begin{equation}\label{eq:xidef}
    \xi_2\equiv R_{\rm E} \frac{\alpha''_{\rm E}}{1-\kappa_{\rm E}},
\end{equation}
where $\kappa_{\rm E}$ is the mean convergence and $\alpha''_{\rm E}$ is the second derivative of the deflection, where the "$E$" subscript refers to the evaluation at $r=R_{\rm E}$. The "2"  subscript on $\xi_2$ refers to being derived from the second order term in the expansion, which we share in Appendix \ref{sec:taylor}. The analogous quantity derived by \citet{Sonnenfeld17} and shown in terms of the radial stretch factor by \citet{Birrer21b}, $\xi_{\rm rad}$, is conceptually equivalent with a difference only of a factor of 2,
\begin{equation}\label{eq:xi_otherdef}
    \xi_{\rm rad} \equiv R_{\rm E} \frac{\psi'''_{\rm E}}{1-\psi''_{\rm E}} = R_{\rm E} \frac{\alpha''_{\rm E}}{1-\alpha'_{\rm E}} = \frac{1}{2}\xi_2,
\end{equation}
noting that $\alpha=\psi'$ and using for a circular lens \citep{Bartelmann10},
\begin{equation} \label{eq:kappa_laplacian}
    \kappa=\frac{1}{2} \nabla^2 \psi 
    = \frac{1}{2} \left[\frac{1}{r}\frac{\partial}{\partial r}\left( r \frac{\partial \psi}{\partial r}\right) \right]
    = \frac{1}{2}\left(\psi''+\frac{\alpha}{r}\right).
\end{equation}
Evaluating at the Einstein radius and recognizing that $\alpha(R_{\rm E})=R_{\rm E}$, one can express $\kappa_{\rm E}$ in terms of the local first derivative of deflection,
\begin{equation} \label{eq:kappa_E_identity}
    \kappa=\frac{1}{2}\left(\alpha'_{\rm E}+1\right),
\end{equation}
reconciling the two definitions of $\xi$ by inserting $\kappa_{\rm E}$ into Eq. \ref{eq:xidef}.

If one mass model is within an MST of another, they will match the same $R_{\rm E}$ and $\xi_2$. If the fit model is a power law, the $\xi_2$ constraint dictates the radial slope of the profile: $\gamma=2+\xi_2/2$. The convergence of the power law can easily be derived as
\begin{equation} \label{eq:kappaE_PL}
\kappa_{\rm E, PL}=\frac{2-\xi_2}{4}.
\end{equation}
The accuracy of the Fermat potential, 
\begin{equation}
    \tau=\frac{(\theta-\beta)^2}{2}-\psi(\theta),
\end{equation} can be shown to be proportional to $(1-\kappa_{\rm E})$ \citep{Kochanek02}. While Fermat potential is somewhat abstract, the observable quantity is the relative time delay between images $\Delta t$, which is given by the Fermat potential, scaled by cosmological distances such that $\Delta t = D_{\Delta t} c^{-1} \Delta \tau \propto H_0^{-1} \Delta \tau$. As such, errors in the Fermat potential are compensated by errors in the recovered value of $H_0$, such that the observable $\Delta t$ is correctly matched. Therefore, one can use $H_0$ as a proxy to represent the accuracy of the Fermat potential, which we elect to use since it is a more intuitive quantity with one specific value for a given model (rather than for each pair of images, which is more cumbersome to express). Through this relation, the expected fractional error on $H_0$ can be expressed using a comparison between the true convergence (which is generally unknown except in simulated lenses) and that of the power-law model. This error is equivalent to the MST parameter $\lambda$ if the difference between the profiles corresponds to an MST:
\begin{equation}\label{eq:h_kappa_scale}
    \frac{1-\kappa_{\rm E, PL}}{1-\kappa_{\rm E, true}} = \frac{H_{0, \rm PL}}{H_{0, \rm true}}= \lambda.
\end{equation}
Finally, one can estimate the width of the recovered $H_0$ posterior, $\Delta H_{0, \rm PL}$, through a propagation of errors on $\xi_2$: 
\begin{equation} \label{eq:errorprop}
    \frac{\Delta H_{0, \rm PL}}{H_{0, \rm true}}=\lambda \frac{\Delta H_{0, \rm PL}}{H_{0, \rm PL}} = \lambda \frac{\Delta\left(1-\frac{2-\xi_2}{4}\right)}{1-\frac{2-\xi_2}{4}} = \lambda
    \frac{\Delta\xi_2}{2+\xi_2} \simeq \lambda \frac{\Delta\xi_2}{2},
\end{equation}
where we have used $\xi_2\sim0\ll2$, noting that galaxy-scale lens systems have mass profile slopes near isothermal, that is, $\gamma=2$ \citep[e.g.,][]{vandeVen10, Auger10, Shajib21}. For example, to reach a 1\% accuracy on $H_0$, the 
 target precision for a population of systems, $\xi_2$ must be known to within an absolute error $\sim0.02$ (assuming $\lambda \sim 1$). Since $\xi_2 \sim 0$, this description using absolute error in $\xi_2$ is more robust than a description using relative error \citep[such as that of][]{Kochanek21}. 

In practice, the utility of $\xi_2$ has been considered for two main purposes. First, because it is considered to be model-independent, a measurement of $\xi_2$ from one model could be used to help guide another model in the optimization process. One example of this idea was demonstrated by \citet{Shajib21}, who included lensing information in a kinematics analysis by folding the posterior distributions for $R_{\rm E}$ and $\xi_2$ into the Bayesian framework, rather than jointly modeling the lensing and kinematic parameters together, thereby under the implicit assumption that the stars and DM in the composite lens share the single ellipticity of the lens model. Secondly, $\xi_2$ has been considered for its application on the simulation side for systematics testing of time-delay cosmography. In many cases one wishes to consider the case where the true lens is a more complex mass distribution than the model, and evaluate the errors introduced by an overly simplistic lens model \citep[e.g.,][]{Cao22,VandeVyvere22a,VandeVyvere22b}. Because the process of creating and fitting mock images is slow, it would be expedient to be able to estimate what the recovered $H_0$ would be from a given model directly from the mass distribution \citep[see][]{Xu16, Tagore18, Gomer20}. We have considered using $\xi_2$ for this purpose (predicting the would-be PEMD fit and using Eq. \ref{eq:h_kappa_scale}), which would enable one to create much larger samples of lenses for systematics checks. Our attempt to confirm this method using the NFWp mocks in TDCVIII did not match our expectation to the desired precision (approximately 3\% discrepancy on $H_0$, see Sect. \ref{sec:setup}). Several explanations of this mismatch are possible, such as a possible bias in the estimate of $\kappa_{\rm E}$, or higher order terms being required in the Taylor expansion used to define $\xi_2$ (see Appendix \ref{sec:appendix_systematics}). In the next section, we show that the main driver of the mismatch unveiled in TDCVIII is directly attributable to the implicit azimuthal structure introduced through the NFWp profile.

\section{Comparing fits to both NFW parameterizations}\label{sec:setup}
The experiment in this work is designed to probe the effect of the choice of NFW ellipticity parameterization in the context of galaxy-scale strong lensing.
The NFW component represents the dark matter of a galaxy, but since no galaxy is purely dark matter, the impact of the NFW parameterization should be studied in the framework of a composite model with both baryon and dark matter mass components. TDCVIII created a population of mock systems analogous to the TDCOSMO lens population using a Chameleon profile \citep{Dutton14} for the baryon component and a NFWp profile for the dark matter component, constructed to have stellar and dark matter mass distributions typical of real lenses. 
The Chameleon profile has a convergence given as
\begin{equation}\label{eq:chameleon}
\begin{aligned}
    \kappa_{\rm Cham}=\frac{A_0}{(1+q_\kappa)}\left[\frac{1}{\sqrt{r_{\rm ell}^2+4w_c^2/(1+q_\kappa)^2}}\right. \\
   \left. -\frac{1}{\sqrt{r_{\rm ell}^2+4w_t^2/(1+q_\kappa)^2}}\right],
\end{aligned}
\end{equation}
with parameters $w_c$ and $w_t$ with $w_t>w_c$ where $A_0$ sets the mass scale at zero radius. and is tailored to closely mimic a S\'ersic profile with a given $R_{\rm Sersic}$ and $n_{\rm Sersic}$, but with the benefit that its lensing potential is analytically expressible.
We create 20 mocks following the same strategy as TDCVIII, but we replace the NFW component first with an NFWp component and later with an NFWm component. We note that the ellipticities of the convergence of the Chameleon and NFW components were not matched in TDCVIII, so we recreate the NFWp mocks using the ellipticity matching procedure discussed in Appendix \ref{sec:ellipticity_matching}. This population of mock lenses and corresponding fits provides an excellent laboratory to explore the role of the implicit azimuthal structure embedded within the NFW parameterization. 

With fits to both populations, we can compare the resulting values of $H_0$ and $\xi_2$ to determine if the parameterization plays a relevant role for time delay cosmography. In both cases, the composite profile will not recover the fiducial value of $H_0$ due to the MST, but if the MST is the only effect at play, the two should recover the same values of $H_0$ and $\xi_2$, biased from the fiducial $H_0$ by the amount predicted by Eq. \ref{eq:h_kappa_scale}. We will show that the implicit azimuthal structure in the potential-based parameterization causes a deviation from this prediction, inconsistent with an MST.

An alternative design for a similar experiment is to use a Chameleon+NFW profile as the model to fit a lens, and to test if the NFW parameterization affects the recovery of $H_0$ and other parameters. Under an ideal setup, such an experiment would be more akin to how time-delay cosmography modeling is implemented and able to more directly attack the question of whether or not the NFW parameterization introduces a bias in these conditions. However, such a setup introduces complexities in that the input azimuthal structure is more difficult to describe and control. While it is possible to create an input population with a reasonable approximation of the azimuthal structure of real systems \citep[see e.g.,][]{VandeVyvere22a,VandeVyvere22b},
we instead elect for this more controlled experiment where lensing degeneracies can be more directly probed because the azimuthal structure can be traced exclusively to the NFW parameterization, with the acknowledgement that this setup is not directly equivalent to the common practice of using an NFW profile to fit a mock.

\subsection{Experiment specifics}\label{ssec:experiment}

We use the same input systems as TDCVIII, which were constructed to match the observed lens population. A large number of two-component profiles were synthesized, and then a subset was selected which matched several observable quantities such as the ellipticities, Einstein radii, and effective half-light radii of real systems. By probing this population of parameters, we ensure that the results of this experiment hold across a range of realistic lenses. We share the parameters for our mock lenses in Table \ref{table:lens_params}. For more details, see \cite{Gomer22}. The result is a set of input two-component profiles with a realistic distribution of parameters, including ellipticities ranging roughly from input $q=0.6$ to $q=0.8$. TDCVIII simply used this input $q$ for both components, resulting in a mismatch between the $q_\kappa$ of the Chameleon stellar component and the $q_\psi$ of the NFWp dark matter component. We instead rescale the $q_\psi$ of the NFWp component according to Eq. \ref{eq:q_kappa} to match the $q_\kappa$ between the two components. We synthesize 20 mock lens images using both the NFWp and NFWm parameterizations.
\begin{table*}
 \centering
    \begin{tabular}{c c c c c c | c c c }

        \multicolumn{6}{c}{Input lens profile parameters} & \multicolumn{3}{c}{Comparison quantities}\\
        $q_\kappa$ & $w_c$ [\arcsec] & $w_t$ [\arcsec] & $A_0$ [\arcsec] & $r_s$ [\arcsec] & $\rho_s$ [M$_{\odot}$ kpc$^{-3}$] & $R_{\rm Sersic}$ [\arcsec] & $R_{E}$ [\arcsec] & NFW $c$\\
        \hline
        0.629 & 0.0164 & 1.89 & 1.72 & 11.28 & $4.21\times10^6$ & 1.54 & 1.87 & 7.74\\
        0.655 & 0.0200 & 2.30 & 1.44 & 7.76 & $9.87\times10^6$& 1.87 & 1.91 & 11.0\\
        0.733 & 0.0139 & 1.60 & 1.30 & 4.51 & $1.36\times10^7$& 1.31 & 1.37 & 12.5\\
        0.642 & 0.0222 & 2.56 & 1.30 & 6.87 & $6.95\times10^6$& 2.08 & 1.57 & 9.52\\
        {\bf 0.640} & {\bf 0.0182} & {\bf 2.09} & {\bf 2.12} & {\bf 10.62} & $\bf 2.42\times10^6$& {\bf 1.70} & {\bf 1.96} & {\bf 6.12}\\
        0.826 & 0.0193 & 2.22 & 1.60 & 5.07 & $1.04\times10^7$& 1.81 & 1.58 & 11.2\\
        0.641 & 0.0148 & 1.71 & 1.38 & 11.95 & $3.00\times10^6$& 1.39 & 1.48 & 6.71\\
        0.676 & 0.0135 & 1.56 & 1.40 & 4.31 & $1.67\times10^7$& 1.27 & 1.50 & 13.6\\
        0.644 & 0.0173 & 1.99 & 1.76 & 13.79 & $2.40\times10^6$& 1.62 & 1.81 & 6.10\\
        {\bf 0.787} & {\bf 0.0158} & {\bf 1.82} & {\bf 1.32} & {\bf 6.45} & $\bf 6.90\times10^6$& {\bf 1.48} & {\bf 1.35} & {\bf 9.50}\\
        0.786 & 0.0209 & 2.41 & 1.62 & 11.09 & $4.22\times10^6$& 1.96 & 1.78 & 7.75\\
        {\bf 0.729} & {\bf 0.0182} & {\bf 2.10} & {\bf 1.68} & {\bf 10.59} & $\bf 5.00\times10^6$& {\bf 1.71} & {\bf 1.86} & {\bf 8.31}\\
        0.614 & 0.0180 & 2.07 & 1.16 & 6.99 & $8.92\times10^6$& 1.69 & 1.53 & 10.5\\
        0.756 & 0.0204 & 2.35 & 1.62 & 8.37 & $6.30\times10^6$& 1.91 & 1.78 & 9.14\\
        0.705 & 0.0156 & 1.79 & 2.22 & 8.51 & $3.18\times10^6$& 1.46 & 1.86 & 6.88\\
        0.663 & 0.0221 & 2.55 & 1.41 & 6.15 & $7.17\times10^6$& 2.08 & 1.59 & 9.64\\
        0.669 & 0.0202 & 2.32 & 1.23 & 6.14 & $1.26\times10^7$& 1.89 & 1.65 & 12.1\\
        0.625 & 0.0181 & 2.08 & 1.73 & 5.53 & $9.82\times10^6$& 1.69 & 1.83 & 11.0\\
        0.702 & 0.0221 & 2.55 & 1.43 & 12.86 & $2.43\times10^6$& 2.08 & 1.58 & 6.13\\
        0.811 & 0.0207 & 2.38 & 1.37 & 4.15 & $1.89\times10^7$& 1.94 & 1.57 & 14.3\\
        \end{tabular}
        \caption{Parameters for the mock lenses used in this work. The parameters left of the solid line are used to construct the mock profiles, while those to the right of the solid line are measured from the profile for easier comparison with real systems. Rows in bold indicate the three systems depicted in Fig. \ref{fig:composite_contours}. All lens light distributions target $n_{\rm Sersic}=4$, and all lenses are placed at $z=0.25$, where the corresponding physical scale is 3.91 kpc per arcsec. NFW concentration $c$ is defined as $r_{200}/r_s$.}
    \label{table:lens_params}
\end{table*}

\begin{figure*}
    \centering 
    \includegraphics[width=0.9\linewidth]{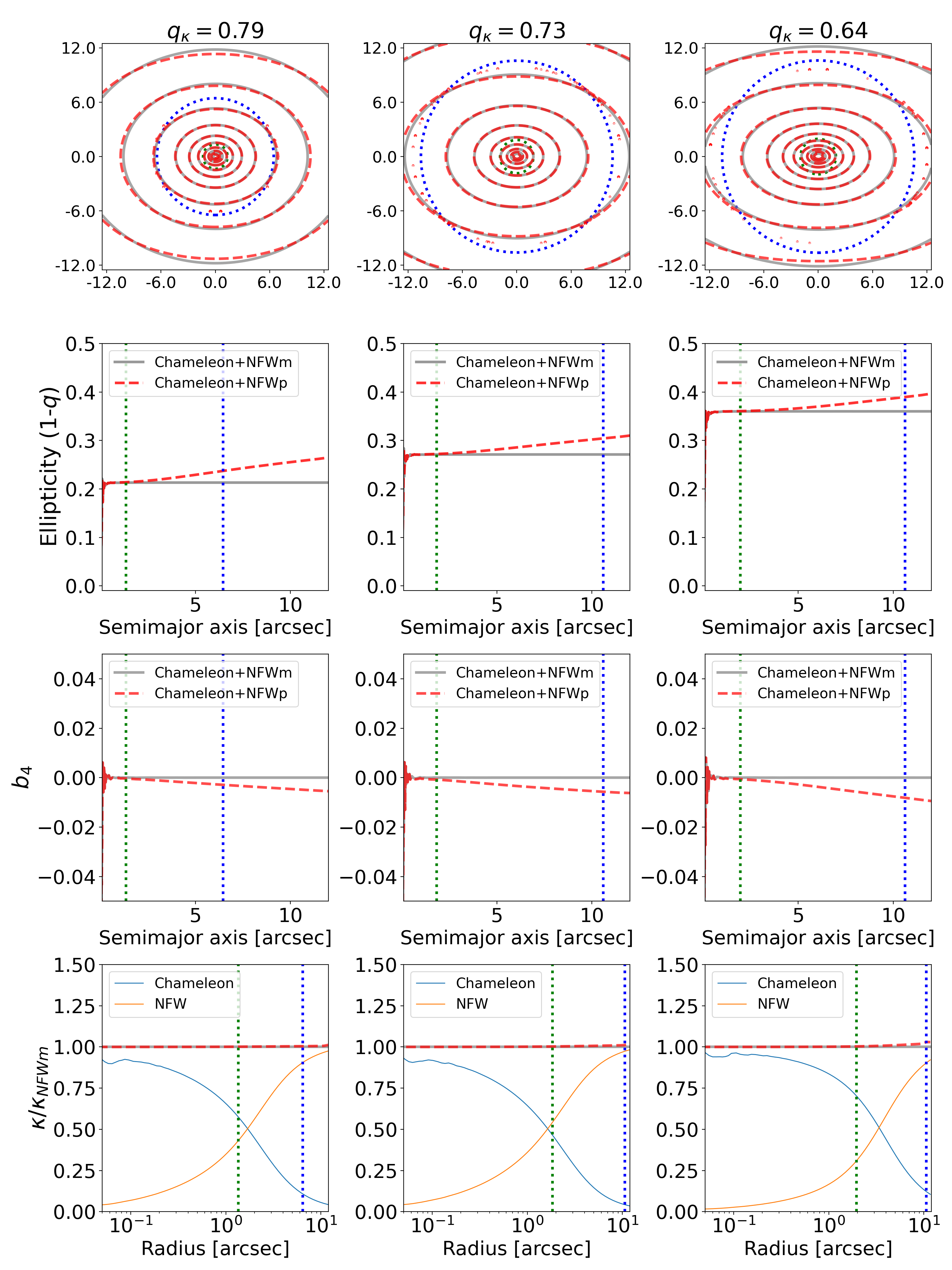}
    \caption{Three composite mass profiles drawn from our sample, with the Einstein radius (green, dotted) and NFW scale radius (blue, dotted) indicated. The implementation of these composite profiles is shown for NFWp (red dashed) and NFWm (gray solid) dark matter profiles. Top row shows isodensity contours; second row shows ellipticity as a function of semimajor axis; third row shows $b_4$ as a function of semimajor axis. The bottom panels show the radial profile over a log scale, defined as the value of $\kappa$ in a circular annulus at a given radius, rather than with respect to semimajor axis, expressed relative to the NFWm radial profile. The composite profiles are plotted as well as both components individually. Units of distance are now in arcseconds.}
    \label{fig:composite_contours}
\end{figure*}

We show three example composite profiles drawn from our population in Fig. \ref{fig:composite_contours} for both the NFWp and NFWm implementations. Note that the field of view differs from Fig. \ref{fig:nfw_contours} to better view the structure at the Einstein radius, which is significantly interior to the NFW scale radius. Inside the Einstein radius, the Chameleon profile dominates, resulting in a nearly perfect elliptical shape. Only in the outer regions does the ellipticity gradient and nonzero boxiness resulting from the NFWp profile become apparent. Radially, the two NFW implementations have essentially identical structure.

Like TDCVIII, we use \texttt{lenstronomy}\footnote{
    \url{https://github.com/lenstronomy/lenstronomy}}
\citep{Birrer18, Birrer21a} to create and fit our mock lenses. Our NFWp mocks are the same as the set used in TDCVIII (specifically the TDCOSMO-like set in that work), except that we rescale the ellipticity so that the mass components match. Our NFWm mocks simply substitute the NFW components for the NFWm parameterization. One last small change we make between our two parameterizations compared is to set the source position relative to the caustic in order to maintain the same image configuration; since the caustic changes slightly between the two parameterizations, this slightly moves the source position. This effect is further quantified in Sec. \ref{ssec:crosssection}.

\begin{figure}
    \centering 
    \includegraphics[width=0.95\linewidth]{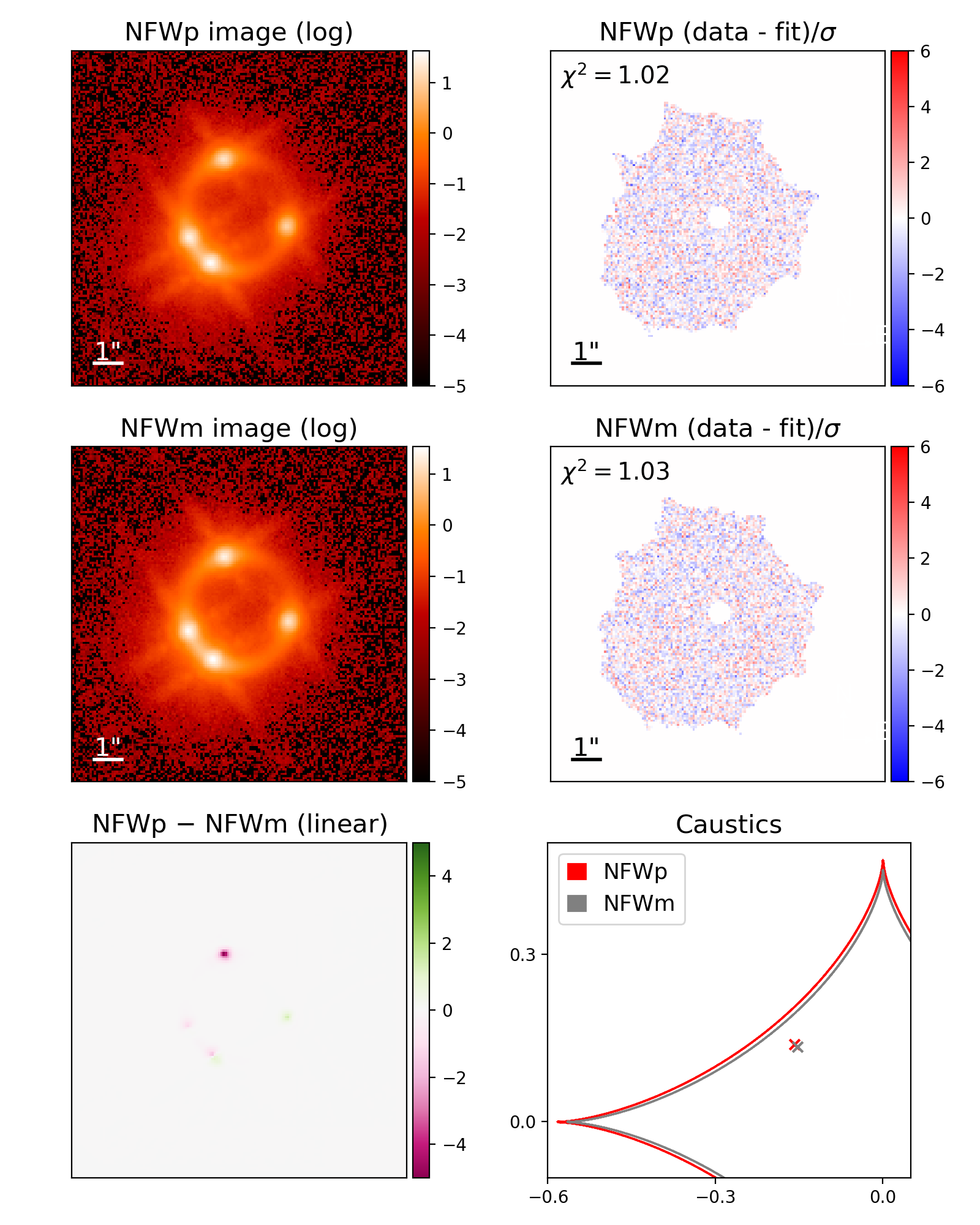}
    \caption{Example image comparison between NFWp and NFWm ellipticity parameterizations (top), with resulting residuals from the PEMD+shear fits (bottom). Caustics for both parameterizations are shown in the bottom right with the source position rescaled as described in the text and  indicated as a cross. This example image comes from a system with moderately large ellipticity with an input axis ratio of $q_\kappa=0.64$.}
    \label{fig:image_comparison}
\end{figure}

We show a comparison between the two mocks for an example image in Figure \ref{fig:image_comparison}. The images are quite similar, with the main differences coming from slightly different point source magnifications with no discernible differences in the arcs. The input Einstein radii and $\xi_2$ values closely match between both parameterizations: the Einstein radii match to within our numerical uncertainty while the input $\xi_2$ values match to better than 0.01 in all cases, with a mean difference of 0.001. 

\subsection{Fit results}

Like TDCVIII, we fit our mock population with a PEMD+shear model. All systems are fit well, with typical residuals comparable to Fig. \ref{fig:image_comparison} (bottom). We find that the azimuthal structures attributable to the NFWp parameterization can be absorbed by the lens model, similar to \citet{VandeVyvere22a,VandeVyvere22b}, who found that nonzero multipole components and ellipticity gradients can often be absorbed by the lens model so long as the deviations from a constant ellipse shape are not too extreme. 

\begin{figure}
    \centering 
    \includegraphics[width=0.95\linewidth]{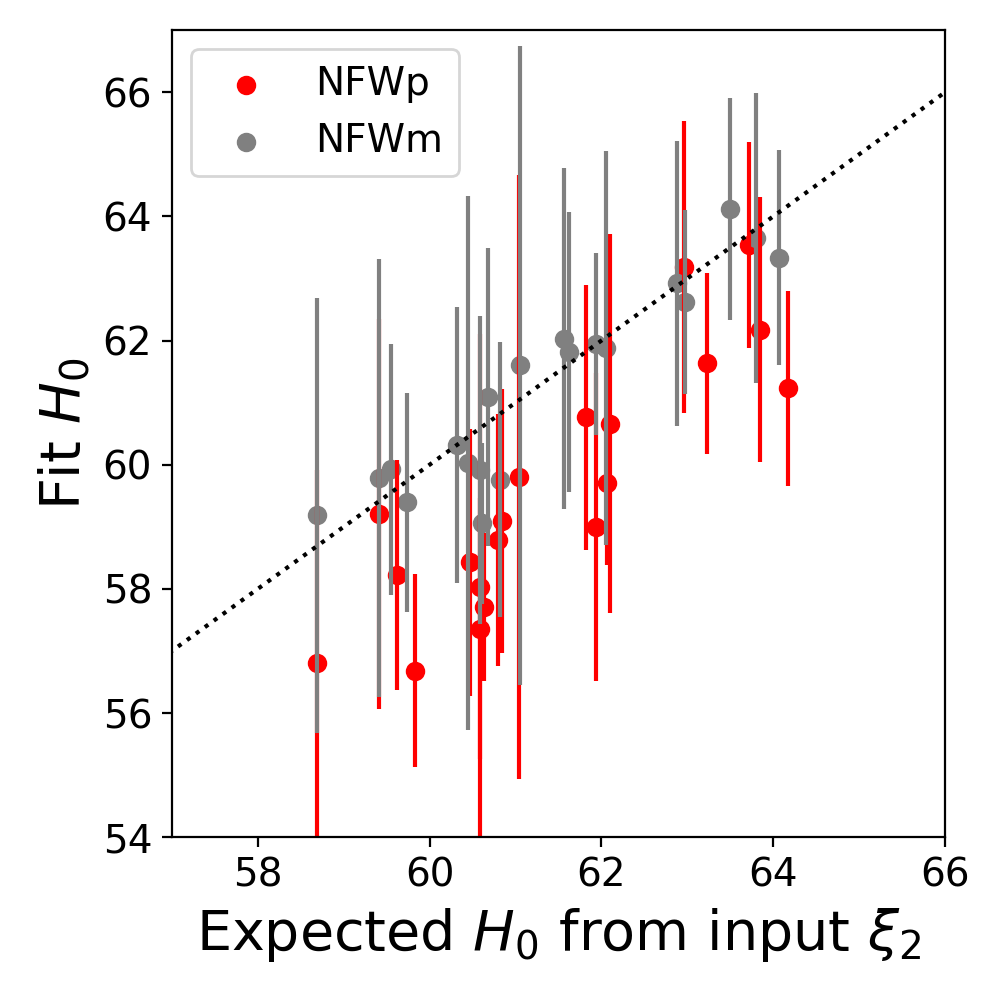}
    \caption{Expected $H_0$ in \Hunits calculated using the $\xi_2$ calculated from the input mass profiles for 20 mock lenses compared to the value of $H_0$ recovered  from a PEMD fit. The dotted line indicates a 1:1 correspondence.}
    \label{fig:input_xi_output_H0}
\end{figure}

If the MST is the only effect at play, the recovered values of $H_0$ should match the predictions according to Eq. \ref{eq:h_kappa_scale} based on the $\xi_2$ of the input mass distribution. We plot the fit values of $H_0$ for both the NFWm population and the NFWp population in Fig. \ref{fig:input_xi_output_H0} alongside the predicted $H_0$. Error bars are estimated via a Markov Chain Monte Carlo (MCMC) estimation using \texttt{emcee} \citep{Goodman10,Foreman-Mackey13}. The fits to the NFWp systems are systematically biased relative to the expectation by approximately $2.5\%$, while the fits to the NFWm systems lie on the expectation line. This result indicates that the Fermat potential recovered by the model is inaccurate by approximately $2.5\%$. This discrepancy appears to be directly caused by the systematic azimuthal structure introduced by the NFWp profile. The effect was originally reported in TDCVIII, although since that work did not match the $q_\kappa$ values between the baryon and DM components and therefore used a more elliptical NFWp component, the magnitude of the effect was slightly larger than seen in this work, quoting a discrepancy of $3\%$.

\subsection{Numerical checks}
\begin{figure}
    \centering 
    \includegraphics[width=0.95\linewidth]{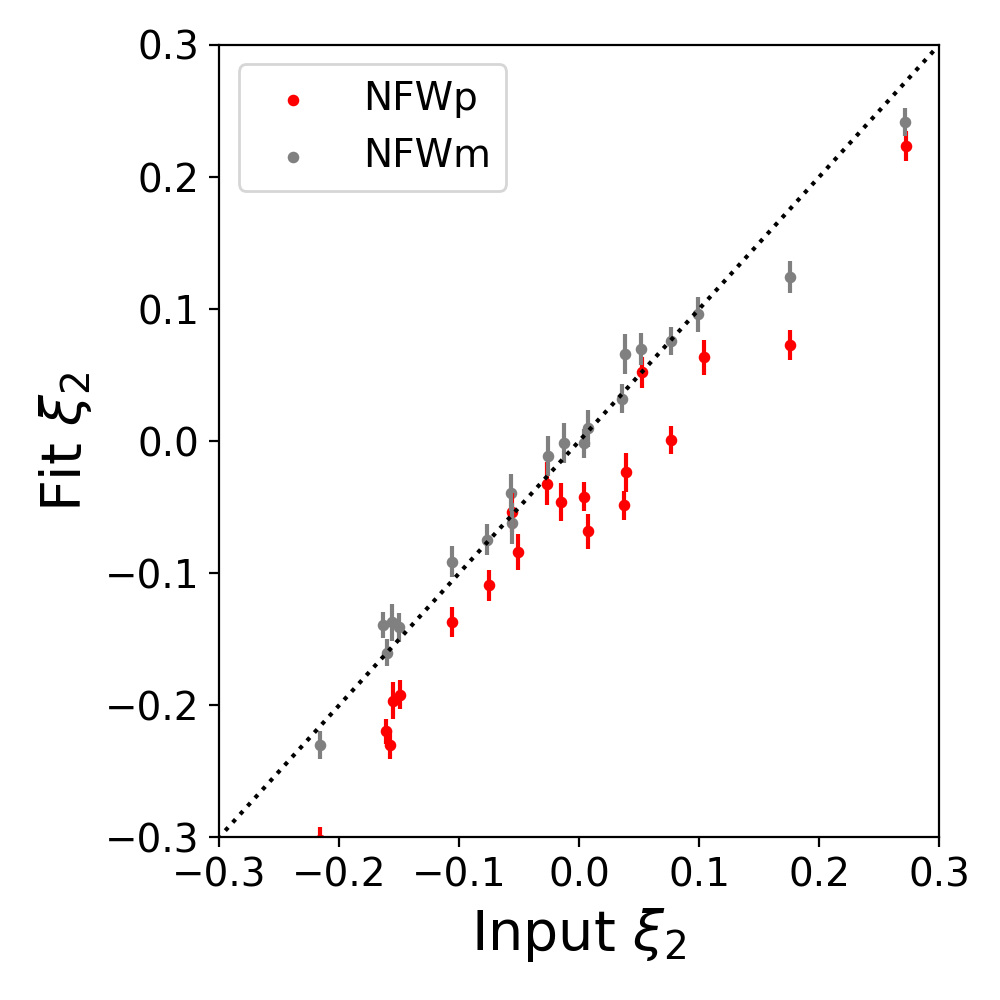}
    \caption{Input $\xi_2$ for the 20 mock lens mass profiles compared to the recovered $\xi_2$ from the PEMD fits, with a 1:1 relation indicated as the dotted line.}
    \label{fig:input_output_xi}
\end{figure}

For completeness, we consider the hypothetical possibility that $\kappa_{\rm E}$ could be numerically biased when comparing one parameterization to another, resulting in the observed $H_0$ discrepancy when Eq. \ref{eq:h_kappa_scale} is applied. As such, we also plot directly the $\xi_2$ values of the PEMD fits compared to those of the input mass distributions in Fig. \ref{fig:input_output_xi}. Again the fits using NFWm parameterization fall tightly on the expected 1:1 line, while those of the NFWp parameterization are systematically biased by approximately 0.05, consistent with Eq. \ref{eq:errorprop}. This result confirms that the discrepancy arises because of implicit azimuthal structure in the NFWp parameterization, rather than numerical effects due to $\kappa_{\rm E}$ estimation.

We have tested several other possibilities to explain this mismatch. Some sources of error we quantified are detailed in Appendix \ref{sec:appendix_systematics}. Namely, we quantify errors due to the truncation of the Taylor expansion used to define $\xi_2$, the effect of numerical errors in $\xi_2$ calculation, and the effect of azimuthal averaging. These effects cannot introduce systematic errors above the percent level, and so we ultimately conclude that the mismatch has been caused by the ellipticity parameterization within the NFW profile.

\section{Discussion} \label{sec:discussion}
We have found that the NFWp parameterization of the NFW profile, when fit with a PEMD+shear model, results in a mismatch between the input $\xi_2$ and that of the fit, ultimately resulting in an $H_0$ which is systematically underpredicted. Theory predicts that the fit value of $\xi_2$ will match the input, since it is the only quantity (along with $R_{\rm E}$) which lensing is able to constrain in an MST-independent manner. So what exactly does this mismatch mean? 

Our interpretation of this result is that the azimuthal structure in the input was not adequately accounted for by the fit model, analogous to the example in \citet{Kochanek21}. The PEMD model, even with external shear, has no ellipticity gradients or nonzero $b_4$ which we now know to be ubiquitous in the NFWp parameterization, shown in Fig. \ref{fig:nfw_contours}. Lacking the capacity to include this azimuthal structure, the PEMD recovered a biased value of $\xi_2$ compared to the value which would have been recovered if the fit had sufficient azimuthal freedom. When the mocks were replaced with those with constant ellipticity, the expected value of $\xi_2$ was recovered. The takeaway is that the expected $\xi_2$ of a fit model and the actual $\xi_2$ of a general mass distribution will not be equivalent unless the azimuthal structure of the model is able to match the truth. 

In other words, this mismatch occurs because the mapping from the input to the PEMD fit cannot be represented by an MST alone, as $\xi_2$ is invariant under the MST by construction. Nonetheless, the resulting image fits are good with no residuals, meaning the input mass model with its {additional azimuthal structure} is degenerate with a PEMD+shear model to within the image noise. We state this explicitly to highlight that the degeneracies at play in this experiment go beyond the MST.

The Source Position Transformation \citep[SPT,][]{Schneider14,Wagner18}, which generalizes the MST to more complex transformations of the radial distribution of the lens, but also yields azimuthal change of the mass \citep{Unruh17}, may appear to be the degeneracy at work in the present experiment. One can show from Eqs. \ref{eq:kappa_laplacian} and \ref{eq:kappa_E_identity} that $\xi_2$ can be expressed as in terms of ratios of derivatives of $\kappa$: 
\begin{equation} \label{eq:SPT_ratio}
\xi_2= R_{\rm E}\frac{\kappa_{\rm E}'}{\kappa_{\rm E}}+1.
\end{equation}
As \cite{Unruh17} show, such ratios of $\kappa$ derivatives are invariant under the MST, but can change under the SPT. Through $\xi_2$, we show that this ratio is unchanged for the fits to the NFWm parameterization, but changed for those of the NFWp parameterization, meaning that the NFWp profiles are not an MST away from the fit, but the transformation is  consistent with an SPT. The SPT is an approximate global degeneracy, and so we check the values of the relative time delays between the input and the PEMD fit. Under an MST, all three time delays are rescaled by the same constant $\lambda$, but under an SPT, the three relative delays are not scaled by the same constant value \citep{Wertz2018}. We find exactly this result, that each individual time delay of the fit is off by the same constant factor for the NFWm mocks but each are off by different factors for each delay for the NFWp mocks, supporting the conclusion that the NFWm mocks are an MST away from a PEMD, while the NFWp mocks are an SPT away from a PEMD. However, the SPT is not a complete description of this degeneracy, because the SPT also transforms the source shape, while all mocks in this work are created and fit using a circular source. Therefore we suspect that there is a related degeneracy at work, likely in the form of a "shape degeneracy" as discussed by \citet{Saha06}, who found similar results using pixelated mass profiles with ellipticity gradients, albeit limited to point sources. Like the SPT, this degeneracy does not uniformly affect time delays, consistent with our findings. Although it is difficult to be more quantitative about their exact form and contribution, it is clear that higher-order lensing degeneracies beyond the MST are certainly at play in this work.

It may look surprising that while most of the azimuthal changes in the input profile appear outside the Einstein radius, (i.e., where the NFW profile starts to dominate), the impact on $\xi_2$ and $H_0$ remains noticeable. This result indirectly shows that morphological assumptions on the density profile {\it beyond} the Einstein radius can have a substantial impact on the lensed images.

We find that the discrepancy between the recovered value of $\xi_2$ for the Chameleon+NFWp and Chameleon+NFWm profiles moderately correlates with the input axis ratio, with Pearson correlation $R=-0.54$. The discrepancy also correlates with the recovered value of external shear, with $R=0.70$. We interpret this to mean that the deviation from an MST worsens with ellipticity, and that external shear can help to absorb this more complex degeneracy. These results add to the body of evidence that external shear can sometimes reflect absorbed degeneracies rather than a physical quantity \citep{Etherington2023}.

One open question for lensing theory is how one should describe the effects of these shape degeneracies in order to include them in lens models. It may be possible to construct a quantity which is independent of these degeneracies analogous to the way $\xi_2$ is constructed to be independent of the MST, but we were unable to derive such a quantity in the confines of this work. The problem is that outside the circular limit, Eq. \ref{eq:kappa_laplacian} requires an azimuthal $\partial ^2 \psi / \partial \phi ^2$ term, entangling azimuthal dependence in any global expression of $\xi$. One should instead use a local quantity such as the stretch differentials discussed by \citet{Birrer21b}. The ultimate goal would be to have a description of how this local quantity, integrated over the imaging information, changes with azimuthal structure. General azimuthal structure is difficult to parameterize, but to test the concept, we created an experiment where we introduced a change in the input $q_\kappa$, but left it unmodeled in the fit, then calculated the expected change in the integrated stretch factor, and found that it corresponds to the amount by which the PEMD $\xi$ recovery is biased. This is still an open field of research, but we believe a complete description of lensing degeneracies requires an exploration along these lines. 

This result puts some limitations on the practical applications of $\xi_2$. Firstly, the practice of using the recovered value of $\xi_2$ from a simple model to place a constraint on a more sophisticated model is not precise to more than a few percent in the general case. If the true mass distribution has azimuthal structure, the simple fit will recover a biased value of $\xi_2$ which then places an inaccurate constraint on the more sophisticated model. This practice can only work if the true mass distribution lacks azimuthal structure, in which case the simple fit will recover an unbiased $\xi_2$. Secondly, the use of $\xi_2$ on the simulation side to estimate the systematic effects of simplistic lens models comes with limitations as well, because the value of $\xi_2$ from an input mass distribution describes what would be recovered by a model that shares the same azimuthal complexity as the data. To perform systematic tests for cases where the azimuthal structure of the mock and of the model differs, the need to create and fit mock lens systems cannot be circumvented.

\section{Connections with other works}\label{sec:otherworks}

In this section we discuss the implications of these results on several related fields of study: namely, measurements of ellipticity and lensing cross section (Sect. \ref{ssec:crosssection}), flux ratio anomalies (Sect. \ref{ssec:fluxanomalies}), and $H_0$ determination (Sect. \ref{ssec:h0}).

\subsection{Ellipticity and lensing cross section} \label{ssec:crosssection}

Two-component lens models are often used to provide observational constraints on the dark matter components of lens systems. In a science case where one wishes to know the ellipticities of dark matter mass distributions, modelers are already wary not to conflate the ellipticity ascribed to the lensing potential with that of the mass \citep{Kassiola93, Barkana98, Golse02}. However, when two-component mass models are used to fit lens systems, the NFWp axis ratio $q_\psi$, is sometimes quoted alongside the baryon component $q_\kappa$ with little distinction made between the quantities, 
opening the door for confusion if one were to take the quoted $q$ at face value. Perhaps deemed irrelevant to a given science case, the distinction between the NFWm and NFWp profile has been largely neglected, and hence the azimuthal structure introduced in the the mass distribution even in the physical low-ellipticity regime has been somewhat overlooked. 

When the NFWp parameterization is used, this slightly changes the caustic size, which can play a role in selection effects affecting lensing studies \citep{Baldwin21}. Studies concerning lensing cross section should be wary about the NFWp parameterization artificially increasing the cross section. 
For the example case in Fig. \ref{fig:image_comparison}, we find the caustic size changes by $3\%$ ($6\%$ by area) compared to the NFWm case. 
We note that this example system has an input $q_\kappa$ of 0.64, which is quite elliptical for our sample. As such this change in caustic size is likely somewhere between the average case and the extreme case. Finally, we also note that quantities defined using 1D profiles, such as the dark matter fraction, are unchanged by the choice of NFW parametrization in the potential or in the mass.

Through the rescaling of ellipticity implemented in this work in Appendix \ref{sec:ellipticity_matching}, it is now relatively simple to convert from a quoted NFWp axis ratio result to what the corresponding mass axis ratio is in the center of the mass distribution. Rather than repeating previous work, this conversion may suffice depending on one's intended scientific application.

\subsection{Flux ratio anomalies} \label{ssec:fluxanomalies}
A second important consideration is the consequences of this work on lensing studies involving calculations of image flux ratios. For cusp-configuration systems, in which the three coalescing images of the cusp each have a signed magnification $\mu$, one can define \citep[e.g.][]{Keeton03} 
\begin{equation}\label{eq:Rcusp}
R_{\rm cusp} = \frac{ \mu_1+\mu_2+\mu_3}{|\mu_1|+|\mu_2|+|\mu_3|},
\end{equation} 
which approaches zero as the source approaches the caustic. Similarly for the two coalescing images in fold systems, 
\begin{equation}\label{eq:Rfold}
R_{\rm fold} = \frac{ \mu_{\rm min}+\mu_{\rm saddle}}{|\mu_{\rm min}|+|\mu_{\rm saddle}|},
\end{equation} 
which also approaches zero as the source approaches the caustic. Deviations from these theoretical cusp and fold relations can be used to diagnose substructure within a lens mass, which have been interpreted as dark matter subhalos \citep[e.g.,][]{McKean07,MacLeod13,Nierenberg14} or as evidence of galaxy group effects or more complex macro-scale mass distributions \citep{Xu15}. 

\begin{figure}
    \centering 
    \includegraphics[width=0.95\linewidth]{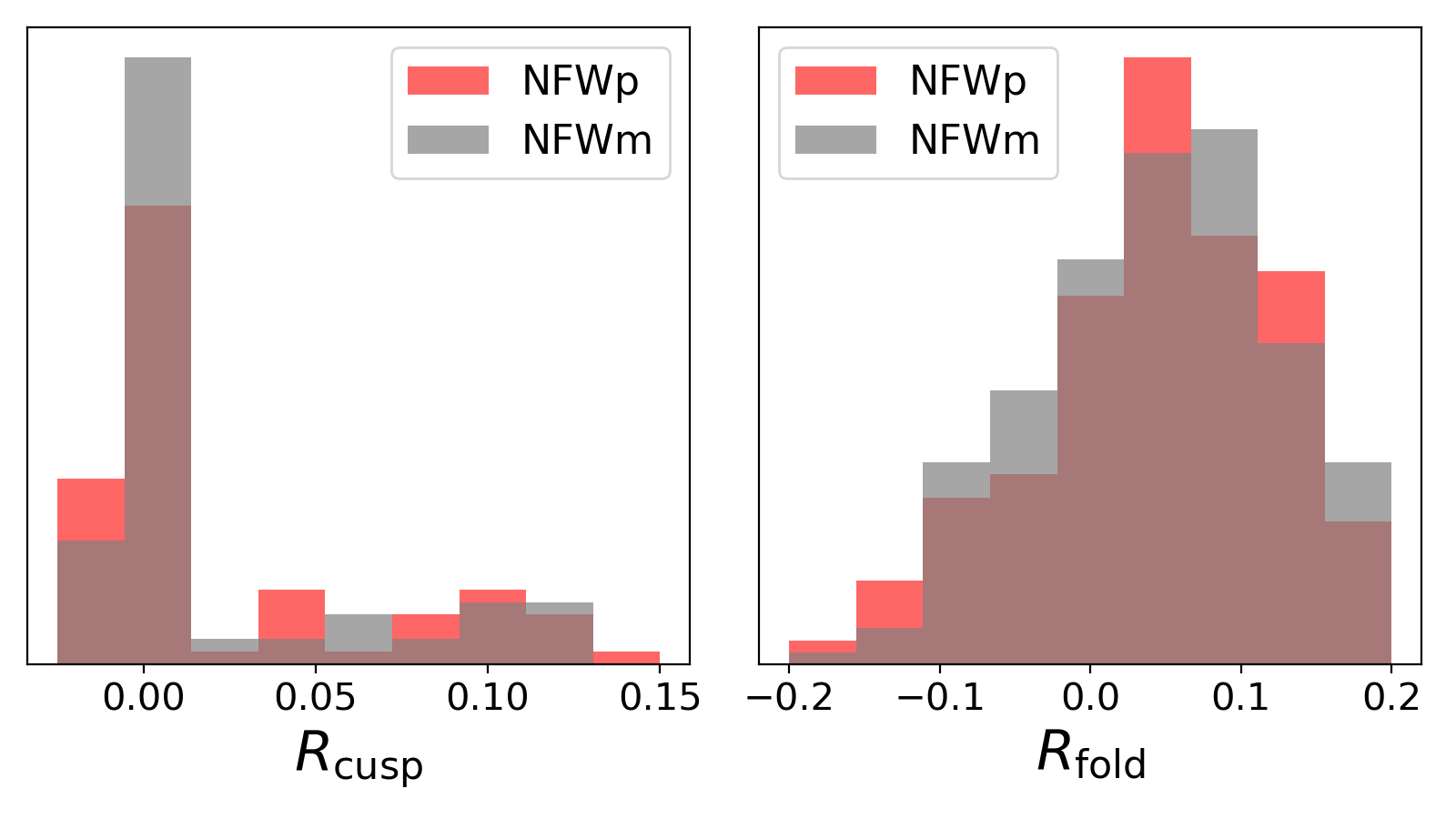}
    \caption{Distributions for $R_{\rm cusp}$ (left) and $R_{\rm fold}$ (right) for many realized source positions of an example mock.}
    \label{fig:RcuspRfold}
\end{figure}

Considerate of this, we evaluate if the distribution of $R_{\rm cusp}$ and $R_{\rm fold}$ would be changed by the NFW parameterization, and so we check this distribution for one of our systems by generating 500 sources for the caustics in Fig. \ref{fig:image_comparison}. Lensing these sources, we designate systems as folds or cusps according to the criteria of \citet{Keeton05}, based on how many images lie within $\sim1 R_{\rm E}$ of one another. We plot the resulting $R_{\rm cusp}$ and $R_{\rm fold}$ distributions in Fig. \ref{fig:RcuspRfold}. We find that the distribution of these flux ratios does not significantly change between the two NFW parameterizations. Comparing the distributions using a Kolmogorov-Smirnov test, we find p-values of 0.16 for $R_{\rm cusp}$ and 0.72 for $R_{\rm fold}$, far from the threshold for similarity typically set at $p<0.05$. As such, we conclude that the NFW parameterization does not directly impact such studies. 

\subsection{$H_0$ determination} \label{ssec:h0}
Because we used NFW profiles in the input mass, rather than fitting a system using an NFW profile, the experiment in this work cannot prove whether or not this effect biases $H_0$ in real systems, but only draws attention to the fact that the NFWp and NFWm make different assumptions. Nonetheless, we believe that the lessons learned in this work regarding azimuthal structures may have some implications for this common use of NFW profiles.

When real systems are modeled, the true mass distribution is unknown. Because the NFWp and NFWm parameterizations give different azimuthal prescriptions, a modeler must decide which description of ellipticity they wish to assume. After all, ellipticity gradients and nonzero $b_4$ exist within real galaxies, and so a modeler may want to select a model which includes such azimuthal structure. However, we note that NFWp azimuthal structure is not the same as what is found in real systems, as it systematically increases ellipticity and boxiness with radius as a nonphysical consequence of the lensing potential, rather than representing the physical tendency of galaxies to have ellipticity which can increase or decrease with radius and structures which can be disky as well as boxy. \citet{VandeVyvere22a, VandeVyvere22b} studied the effect of such azimuthal structures in lens models, broadly concluding that individual lenses can result in biased parameter recoveries while the population averages out to an unbiased determination of $H_0$. Modelers should therefore be wary about the azimuthal structure introduced by the NFWp model; since this effect always applies in the same direction, it will not average out over the population of lenses. If the intention is that a mass profile has a constant elliptical shape, using the NFWp parameterization is not consistent with this assumption and would introduce a systematic effect in the modeling.

However, this is not the whole story when it comes to modeling the azimuthal structure of real systems. When systems are modeled with a composite profile, they are done so using more complex models than we have used here. Centroid positions, position angles, and ellipticities may be allowed to be offset between the two components \citep[e.g.,][]{Rusu20,Shajib22b}. It is entirely possible for choices in this parameter space to compensate for additional azimuthal structure in the NFWp component. As a simple example, using a circular NFW component would result in an ellipticity gradient going from elliptical in the center to circular in the outer regions, in the opposite direction as the NFWp ellipticity gradient, perhaps negating or reversing its effect. 

In addition, nearby perturber galaxies are included in the model \citep{Wong18,Birrer19,Shajib20} and pixelated corrections to the lensing potential can be implemented \citep{Suyu10}. These numerous considerations add considerable azimuthal freedom to the model. This freedom may be sufficient to describe the true mass model, in which case $H_0$ would not be biased beyond the MST. This possibility has not been directly tested, but the fact that TDCOSMO recovers consistent $H_0$ values between their power law models and composite models supports this hypothesis \citep[see Fig. 6 of][]{Millon20}. We do however note that in such a case the particular values of the individual components may not correspond to the true mass distributions, instead seeking a compromise which compensates for the NFWp gradients.

Furthermore, the present work does not include stellar kinematic constraints, which play a vital role in breaking lensing degeneracies \citep{Birrer20,Yildirim23, Shajib23}. Finally, we note any implications of the systematic effects discussed in this work would only apply to NFW profiles and therefore only to the composite fits of TDCOSMO, having no bearing on the power-law fits also adopted by TDCOSMO.

\section{Conclusion} \label{sec:conclusion}

The NFW profile is a key ingredient of realistic models of galaxies, but cannot be described analytically in the elliptical case. For lensing applications, the ellipticity can either be added in the potential (NFWp parameterization) or in the mass via an approximated profile (NFWm parameterization). When ellipticity is introduced in the potential, it introduces azimuthal structure in the form of ellipticity gradients and nonzero boxiness in the mass distribution. We create two populations of composite mocks using each parameterization and fit them with a PEMD+shear model. The mocks created using the potential-based parameterization result in fits which recover biased values of $H_0$ relative to those from the mocks without this introduced azimuthal structure. This result has several consequences:

\begin{itemize}
    \item The use of a potential-based parameterization of ellipticity introduces ubiquitous azimuthal structure in $\kappa$ in the form of ellipticity gradients and nonzero $b_4$, even for low values of ellipticity when the distribution is not yet dumbbell-shaped. We note that presence of artificial variations of ellipticities can be mistakenly absorbed by shear \citep{VandeVyvere22b, Etherington2023}. We advise lens modelers who wish to assume an azimuthal shape with constant ellipticity to use the NFWm parameterization (which may be based on CSEs as we have used in this work or another formulation that keeps ellipticity constant with radius) in order to be consistent with this assumption. Azimuthal structure can still be implemented through multiple mass components with differing ellipticities or position angles, but it would be done so explicitly rather than unintentionally.
    \item Our fits to mocks with both NFW parameterizations resulted in values of $H_0$, which were discrepant with one another by $2.5\%$ (systematic), indicating an inadequacy of the model to capture the true Fermat potential at this level. However, we cannot claim that the practice of fitting mocks with NFW models necessarily introduces a bias on $H_0$ at the same level due to the additional azimuthal freedom of TDCOSMO-like models, which may compensate for this effect.
    \item The MST-independent quantity $\xi_2$ is an accurate predictor of the recovered mass model in the case where the input and the fit have the same azimuthal prescription. However, when the azimuthal structure in the input mock is not captured by the PEMD model, the value of $\xi_2$ is biased, indicating that the mapping is not a simple MST, and may in fact be a more general SPT or even a shape degeneracy. Various tests and subtleties in the possible ways to calculate $\xi_2$ are described in Appendix \ref{sec:appendix_systematics}.
    \item The bias introduced by the NFWp parameterization is mostly caused by deviation from elliptical isodensity contours that take place {\it beyond} the Einstein radius. This indirectly shows that morphological assumptions motivated solely by the shape of the lensing galaxy interior to the Einstein radius and/or a generic absence of ellipticity of the NFW component may introduce a bias as large as several percent in some lensing-inferred quantities such as $H_0$, $q$, or the external shear magnitude. 
    \item As we have shown that $\xi_2$ differs between two models with different azimuthal structure, it can only be used to convert one radial profile to another by keeping the same model assumption on the azimuthal structure. The use of $\xi_2$ derived from PEMD modeling has been considered as proxy for constraining other models, such as a composite model, without directly optimizing the model on the lensed images. As an early example of such an application, \citet{Shajib21} used the $\xi_2$ from a PEMD lens model to constrain a composite model's radial profile in dynamical modeling, although in this example the model was spherical and as such the treatment of azimuthal structure is irrelevant. More generally, the accuracy of this procedure is limited to the amount by which the original power-law lens model is able to capture azimuthal structure: in our case having an inaccuracy of approximately 0.05 on $\xi_2$.  If high accuracy is required, we caution against the use of $\xi_2$ as a constraining diagnostic when comparing models with differing azimuthal structure. 
    \item For a test case with significant ellipticity (input $q=0.64$), the introduced azimuthal structure also changes the cross section for quad lenses by $\sim6\%$. This is particularly relevant for understanding the selection function of lensed systems in existing and upcoming large surveys such as Euclid \citep[e.g.,][]{Sonnenfeld23}. 
    \item Flux ratios remain broadly unaffected by the choice of NFW parameterization of the macro model of the lens. 
\end{itemize}

While the true mass distributions of lenses are not exactly known, it is important to quantify the effects of implicit assumptions inherent in the choice of lens model. The exact prescription of the NFW ellipticity is one such assumption which we show can have an effect if high accuracy is required.

\begin{acknowledgement}

In addition to those mentioned in the text, this work uses the following Python packages: Python \citep{Python1,Python2}, Astropy \citep{astropy:2013,astropy:2018}, Numpy \citep{Numpy}, Scipy \citep{scipy}, Matplotlib \citep{Matplotlib}, Pandas \citep{pandas1,pandas2}, and Seaborn \citep{seaborn}.

We thank the referee, whose insights regarding the ellipticity of the NFWp mass distribution are reflected in Appendix \ref{sec:ellipticity_matching}, improving the experiment design of this work.

This project has received funding from the European Research Council (ERC) under the European Union’s Horizon 2020 research and innovation program (grant agreement No 787886). 
Support for this work was provided by NASA through the NASA Hubble Fellowship grant HST-HF2-51492 awarded to AJS by the Space Telescope Science Institute, which is operated by the Association of Universities for Research in Astronomy, Inc., for NASA, under contract NAS5-26555.
\end{acknowledgement}

\bibliographystyle{aa} 
\bibliography{biblio}

\appendix

\section{Ellipticity matching to an elliptical potential} \label{sec:ellipticity_matching}

Given that a lensing potential results in elliptical contours, this gives rise to a non-elliptical shape of the convergence profile, including ellipticity gradients. Here we analytically show the origin of these features, as well as analytically calculate the mass axis ratio for an NFW potential in the limit of small radius. Starting with a general potential, consider any elliptical potential of the form 
\begin{equation}
    \psi(x,y) = f(a),
\end{equation}
where 
\begin{equation}
a=\sqrt{x^2+\frac{y^2}{q_\psi^2}}
\end{equation}
is the semimajor axis which captures all the dependence on the $x$ and $y$ coordinates. Note that the elliptical radii discussed in this work, $r_{\rm ell}=a\sqrt{q_\psi}$ and $r_{\epsilon}=r_{\rm ell} \sqrt{2q_\psi/(1+q_\psi^2)}$, both take on the form $g(q_\psi)*a$ and as such capture all spatial dependence in terms of $a$, resulting in an elliptical shape. 
The convergence then takes the form 
\begin{equation}
\begin{split}
    \kappa(x,y)&=\frac{1}{2}\nabla^2\psi(x,y)\\
    &=\frac{1}{2a}\left(1+\frac{1}{q_\psi^2}\right)f'(a) + \frac{1}{2a^2} \left(x^2 +\frac{y^2}{q_\psi^4}\right)\left(f''(a)-\frac{1}{a}f'(a)\right)
\end{split}
\end{equation}
A function of this general form 
\begin{equation}
    \kappa(x,y) = A(a)+ \left(x^2 +\frac{y^2}{q_\psi^4}\right)B(a)
\end{equation}
is no longer a pure function of $a$, now containing additional $x$ and $y$ dependence, ergo no longer having purely elliptical contours. To calculate the axis ratio of this convergence, one can set $x$ or $y$ equal to zero to calculate the value of $\kappa$ along the major axis or minor axis:
\begin{equation}\label{eq:kappa_along_axis}
    \begin{split}
    \kappa(x, y=0)&=\frac{1}{2}\left[\frac{f'(a|a=x)}{x q_\psi^2}+f''(a|a=x)\right] \\
    \kappa(x=0,y)&=\frac{1}{2}\left[\frac{q_\psi}{y}f'\left(a|a=\frac{y}{q_\psi}\right)+\frac{1}{q_\psi^2}f''\left(a|a=\frac{y}{q_\psi}\right)\right].
    \end{split}
\end{equation}
By setting $\kappa(x_c, y=0)=C$ and $\kappa(x=0, y_c)=C$, one can solve for the $x_c$ and $y_c$ values corresponding to a particular isocontour with value $C$. This solution cannot be expressed for a general potential, but one can show that the NFW potential (Eq. \ref{eq:nfw_circ_pot}) can be expressed in the limit of small $a$ as 
\begin{equation}
    \psi_{\rm NFW} \simeq -a^2\log{\frac{a}{2}},
\end{equation}
resulting in invertible expressions for Eqs. \ref{eq:kappa_along_axis}. Solving for $x_c$ and $y_c$ and taking the ratio of $y_c/x_c$ gives the convergence axis ratio of a given contour, which in this limit is independent of the isocontour $C$,
\begin{equation}\label{eq:q_kappa}
    q_{\kappa, a\rightarrow0}= \frac{y_c}{x_c}= q_\psi \exp{\left(\frac{q_\psi^2-1}{q_\psi^2+1}\right)}.
\end{equation}
We use this expression to calculate the input $q_\psi$ for our NFWp profiles, guaranteeing that they have the same $q_\kappa$ as the Chameleon profiles at innermost radii.

\begin{figure}
    \centering 
    \includegraphics[width=0.95\linewidth]{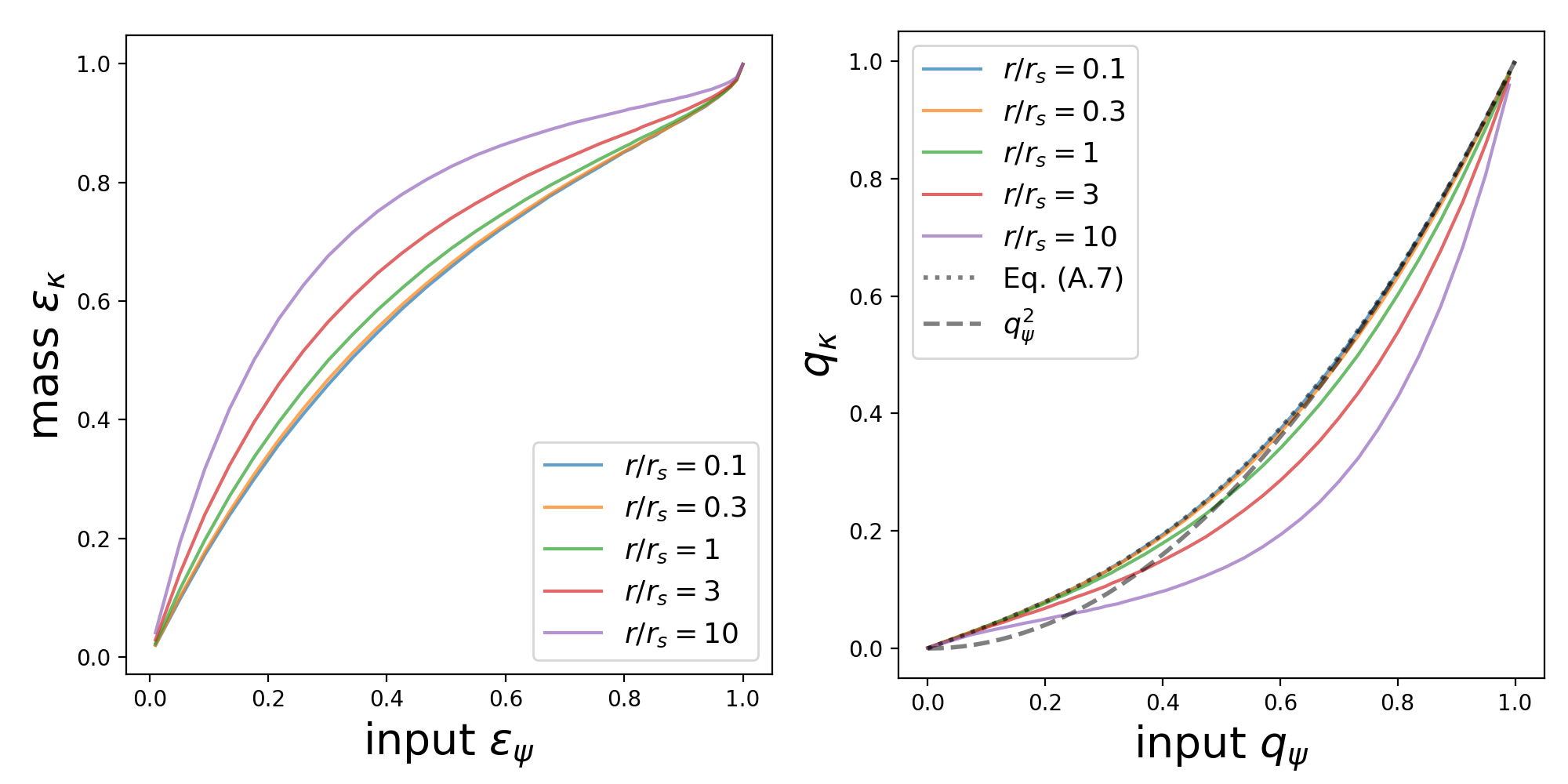}
    \caption{Relationship between the mass ellipticity and potential ellipticity for the NFWp profile. Left: ellipticity for different $r$ values. Right: same relation in terms of axis ratios.}
    \label{fig:recreate_GK}
\end{figure}

To more completely illustrate the relationship between $q_\psi$ and $q_\kappa$, for an NFW profile, we recreate Fig. 2 of \citet{Golse02}, which gives the mass ellipticity as a function of the input potential ellipticity, where $\epsilon=(1-q^2)/(1+q^2)$. The mass axis ratio is calculated numerically by determining the point along the $y$ axis which has the same convergence as a point on the $x$ axis and taking $q_\kappa=y_c/x_c$. We plot this in the left panel of Fig. \ref{fig:recreate_GK}. This figure shows that the relation between the two ellipticities is never equality, and the difference between them changes as a function of $r$: an ellipticity gradient. We also plot the same relation in terms of axis ratios $q_\psi$ and $q_\kappa$ (right panel). Plotted this way, one can see clearly that $q_\kappa$ goes as Eq. \ref{eq:q_kappa} in the limit of small $r$, which is well-approximated as $q_\psi^2$ for $q_\psi>0.6.$ 

\section{Systematics checks}\label{sec:appendix_systematics}
\subsection{Taylor expansion}\label{sec:taylor}

We discuss here the Taylor expansion of \citet{Kochanek20} used to define $\xi_2$, and its accuracy as a function of the limiting order of the expansion.
Expanding deflection for a circular lens for an image near $R_{\rm E}$:
\begin{equation} \label{eq:alpha_exp}
\alpha(r) = R_{\rm E}+\alpha'_{\rm E}(r-R_{\rm E}) + \frac{1}{2}\alpha''_{\rm E}(r-R_{\rm E})^2+\frac{1}{6}\alpha'''_{\rm E}(r-R_{\rm E})^3+... 
\end{equation}
From the lens equation ($\beta=\theta-\alpha$), where $\theta=r$ for a given image, $\beta$ is given to second order in $(r-R_{\rm E})$ as
\begin{equation}\label{eq:beta_exp}
    \beta(r) \simeq -2(\kappa_{\rm E}-1)(r-R_{\rm E}) - \frac{1}{2}\alpha''_{\rm E}(r-R_{\rm E})^2.
\end{equation}
where we have used Eq. \ref{eq:kappa_E_identity}.
Dividing by $(1-\kappa_{\rm E})$,
\begin{equation} \label{eq:2terms}
    \hat{\beta}(r)\equiv \frac{\beta(r)}{1-\kappa_{\rm E}} \simeq 2(r-R_{\rm E}) - \frac{1}{2}\frac{\alpha''_{\rm E}}{1-\kappa_{\rm E}}(r-R_{\rm E})^2.
\end{equation}
One can see by substituting Eqs. \ref{eq:mst} and \ref{eq:mst_beta} that $\hat{\beta}(r)$ is invariant under the MST. As such, the right-hand side of the equation is also MST-invariant. Furthermore, since $R_{\rm E}$ is MST-invariant, the second-order term is MST-invariant as well. From here, $\xi_2$ is defined according to Eq. \ref{eq:xidef} by making the second term unitless via a factor of $R_{\rm E}$.

Let us consider the error associated with the truncation of the Taylor expansion. Using any finite order for a Taylor expansion will introduce some discrepancy from the truth $\Delta \hat{\beta}$:
\begin{equation} \label{eq:beta_error}
    \hat{\beta} = \hat{\beta}_{\rm Taylor} + \Delta \hat{\beta},
\end{equation}
where, for two terms, $\hat{\beta}_{\rm Taylor}$ is defined by Eq. \ref{eq:2terms}. Since $R_{\rm E}$ is measured very precisely, consider that the error due to the truncation of the Taylor expansion will be interpreted as error in $\xi_2$. Rearranging this equation for a second order $\hat{\beta}_{\rm Taylor}$,
\begin{equation} 
    \hat{\beta} = 2(r-R_{\rm E}) - \frac{1}{2 R_{\rm E}} (r-R_{\rm E})^2 \left[ \xi_2-\frac{2 R_{\rm E} }{(r-R_{\rm E})^2} \Delta \hat{\beta} \right].
\end{equation}
We define the error in $\xi_2$ as 
\begin{equation} 
    \Delta \xi_2 = -\frac{2 R_{\rm E} }{(r-R_{\rm E})^2} \Delta \hat{\beta}.
\end{equation}
We construct this quantity this way because we are curious how much the error associated with the truncation of the Taylor expansion can result directly in an error in $\xi_2$. This implicitly assumes that there is no error in $R_{\rm E}$ such that all of the truncation error is applied to $\xi_2.$ 

Though this quantity is defined using  two terms in the Taylor expansion, it can be useful to calculate it using $n$ terms in the expansion for $\hat{\beta}_{\rm trunc}$ to estimate the increased accuracy of a higher order expansion. With three terms, for example, $\Delta \xi_2$ under this construction assumes all error associated with the truncation still applies to $\xi_2$, which is strictly speaking inaccurate because $\xi_3$ should have error associated with it. Therefore, this representation serves as a conservative estimation of the maximum possible error in $\xi_2$, signifying by how much $\xi_2$ would need to change to compensate for the error in the expansion.

\begin{figure}
    \centering 
    \includegraphics[width=0.95\linewidth]{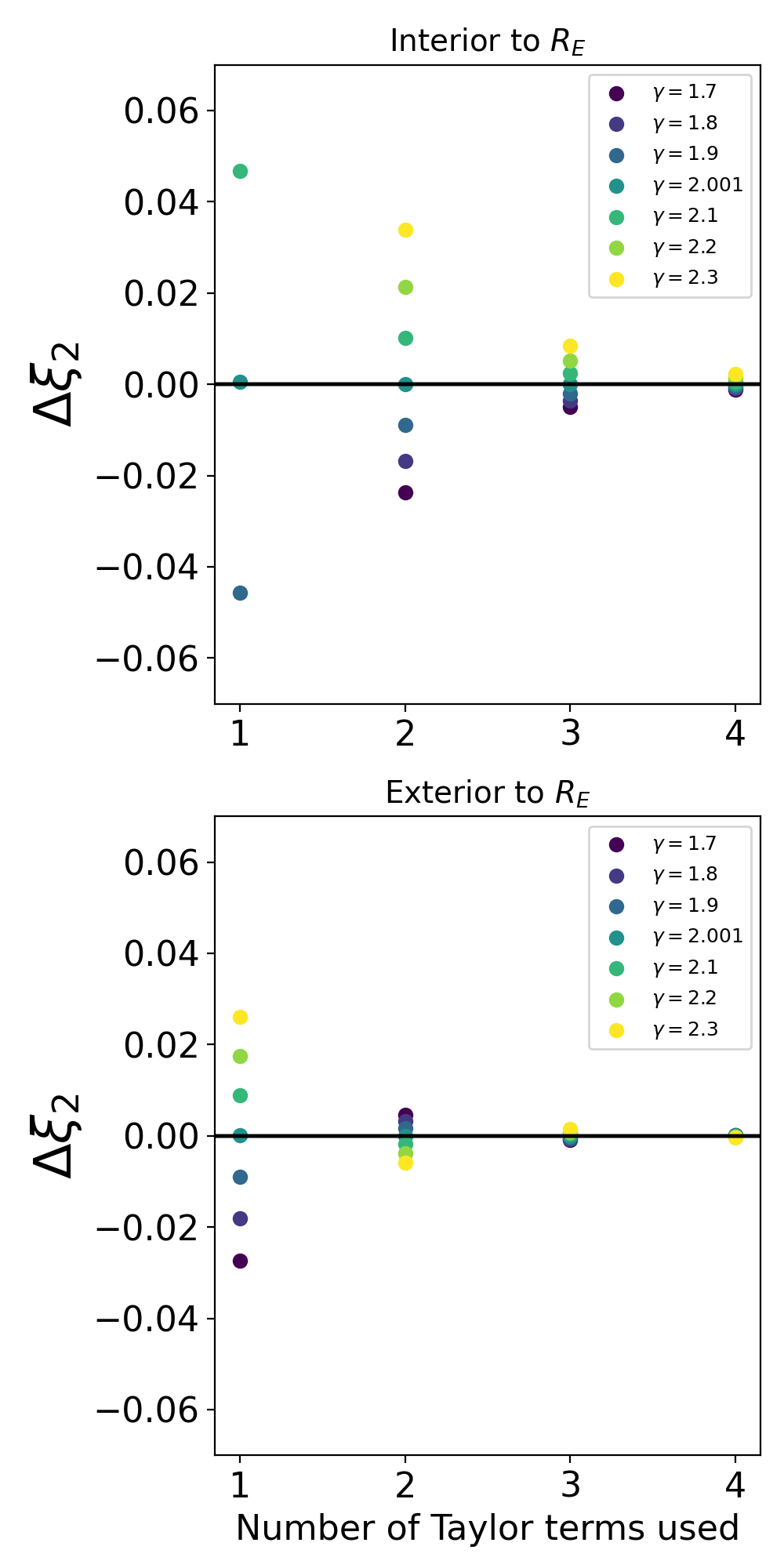}
    \caption{Error resulting from Taylor expansions of deflection of different order, interpreted as error in $\xi_2$, using $|r-R_{\rm E}|/R_{\rm E}=0.3$. In the top panel, $|r|<R_{\rm E}$, while in the bottom panel, $|r|>R_{\rm E}$. The $\xi_2$ formalism is equivalent to using the second order expansion.}
    \label{fig:taylor_xi}
\end{figure}

We plot this quantity in Fig. \ref{fig:taylor_xi} using a power law for which the true $\hat{\beta}$ can be calculated analytically, using several different slopes and several values of $n$. We evaluate this error both in the case where $r=0.7R_{\rm E}$ (interior to $R_{\rm E}$) and when $r=1.3R_{\rm E}$ (exterior to $R_{\rm E}$), such that $|r-R_{\rm E}|/R_{\rm E} = 0.3$ in both cases. We find that for slopes between $\gamma=1.7$ and $\gamma=2.3$, the error is centered on zero with lessening scatter as the number of terms increases. The error is approximately 5 times larger for images interior to $R_{\rm E}$ than those exterior to $R_{\rm E}$, but for $n=2$ the error is always less than 0.04. This makes it unable to explain the systematic difference of $\simeq 0.06$ discussed in this paper. Furthermore, since the error is centered on zero, a population with an average slope of 2 will not have any systematic bias, although a population with a mean slope significantly different than 2 could result in a systematic bias of order 0.02, depending on the slope. 
Using additional Taylor terms decreases this error as expected, but to implement higher order terms in practice would require more careful accounting of errors on $\xi_3$ or $\xi_4$: a complexity which we have neglected. This result is in agreement with that of \citet{Birrer21b}, who quoted better than $1\%$ accuracy on deflection using the second order expansion under similar conditions.

With a maximum error on $H_0 \simeq 1\%$ which is not systematic, the Taylor truncation cannot explain the $3\%$ systematic discrepancy in this work. Furthermore, we find that the width of the region probed by the images does not correlate with the value of the $H_0$ mismatch (as one would expect errors to grow with $|r-R_{\rm E}|$). We therefore conclude that the Taylor truncation error is not the cause of the discrepancy shown in Fig.~\ref{fig:input_output_xi}.

\subsection{Numerical robustness of $\xi_2$} \label{sec:numeric_xi}
As this work considers a comparison between similar values of $\xi_2$, it is important to quantify the effects of numerical precision. We calculate the $\xi_2$ of the input mass distribution in \texttt{lenstronomy}, which in its current implementation does so by sampling a ring of points at the Einstein radius and evaluating the radial derivatives of the lensing potential at each point, then taking an average over the set for a single value of $\xi_2$. We quantify the robustness with respect to the choice of the number of points used to sample the ring. 

We first compare the calculation to the analytical power law case over the range of slope values in this work, which ranges approximately from $\xi_2=-0.3$ to $0.3$, based on Fig.~\ref{fig:input_output_xi}. This corresponds to $\gamma \in [1.85,2.15]$. In the circular case, \texttt{lenstronomy} calculates $\xi_2$ exactly, regardless of the number of points. In the elliptical case (where we set $q=0.6$), we find that the numerical calculation is within 0.002 of the true value of $\xi_2$, with accuracy which improves as we increase the number of points up to 1000 points, where it remains constant at approximately 0.001. Interestingly, we find that for profiles with steeper slopes, the accuracy on $\xi_2$ does not continue to improve with an increased number of points, and has a systematic bias at the <0.001 level. 

We also test this effect on the composite profiles, although we cannot analytically calculate the truth value of $\xi_2$. Testing with the Chameleon+NFWp profiles, we find that the value is not robust (changes by approx. 0.02 or more) with fewer than 1000 points, but the robustness improves with more points. In particular, the change from 3000 points to 10000 points changes the evaluation of $\xi_2$ by 0.002 (median) with 0.004 standard deviation. We consequently chose to use 3000 points for this work. The Chameleon+NFWm profiles are even more robust, with a change of only $4\times10^{-5}$ (median) with 0.0002 standard deviation when increasing from 3000 to 10000 points.

From these tests, we conclude that the numerical calculation of $\xi_2$ is robust to within at worst 0.004 statistical scatter ($0.2\%$ for $H_0$) with a systematic bias likely less than 0.002 ($0.1\%$ for $H_0$). This effect therefore cannot explain the discrepancy between the fitted $\xi_2$ values originating from the two profile parameterizations in this work.

\subsection{Other explored effects }\label{sec:azimuthal}

The derivation of $\xi_2$ comes from the circular limit, and so we consider some of the complexities that arise in the elliptical case. The first main effect we consider is that of propagating the uncertainty on the Einstein radius. The Einstein radius is a circularly averaged quantity and as such we must confirm that this averaging is robust in our non-circular mocks. Like the Einstein radius, $\xi_2$ is circularly averaged, although it may be that the regions where it is most accurately probed are the image locations rather than a uniform circle. As such, the second main effect we consider is the difference between a circularly averaged $\xi_2$ and that of an averaging based on the image positions.

An error in the calculation of the Einstein radius would result in evaluating $\kappa_{\rm E}$ and $\xi_2$ at a different location, which could in principle bias the result. With elliptical distributions, it is important to evaluate the effective Einstein radius, which differs from the normalization value that describes the Einstein radius for circular distributions. We evaluate the effective Einstein radius using a grid and by taking radial steps outward, calculating the circle within which the average density is equal to the critical density. We perform some robustness checks by evaluating the effective Einstein radius for several samples within the MCMC chain of the fit and find that the fit Einstein radius always lies within one radial step of the input, in our case approximately 0.013\arcsec, which we adopt as our uncertainty in $R_{\rm E}$. We then evaluate the change in $\kappa_{\rm E}$ by evaluating the local convergence at this new Einstein radius and find that it changes by approximately 1\% for both the input profile and the fit profile. The worst case scenario would be that the Einstein radius is recovered on the lower end for one and on the higher end for the other, resulting in the ratio in Eq. \ref{eq:h_kappa_scale} being off by approximately 1\%, although this appears equally likely for the NFWp parameterization compared to the NFWm parameterization and as such it is difficult to see how this could create a systematic bias. Similarly, we also evaluate the change in $\xi_2$ of the input profile associated with this change in evaluation radius. For the Chameleon+NFWm, the median change is approximately 0.002, with little spread. For the Chameleon+NFWp, the median change is about half as much, but with significantly more scatter (approximately 0.007). Neither of these effects can explain the observed discrepancy.

Another effect we consider is that $\xi_2$ itself is a circularly averaged property, which makes sense because the profile used to fit the lens to a particular $\xi_2$ is a global description. However, we were curious if there could be an effect due to the local measurements where the lensing information most directly probes, that is, at the image positions. As such, we evaluated $\xi_2$ using the image radial position instead of the Einstein radius for each of the four images, and took the mean of the four evaluations. We also tried taking a weighted average based on the brightness of the images. In either case, the value of $\xi_2$ changes from the traditional circularly averaged calculation by a median of less than 0.005, with spread of less than 0.025, insufficient to explain the discrepancy.

\end{document}